\documentclass[letterpaper, 12pt]{article}[2025/05/19]
\usepackage[english]{babel}
\usepackage{amsfonts,amsmath,amssymb,amsthm,latexsym,amscd,mathrsfs,bbm}
\usepackage{ifthen,cite}
\usepackage[bookmarksnumbered=true]{hyperref}
\usepackage{times,fancyhdr}

\usepackage{amsmath}
\numberwithin{equation}{section}

\usepackage{graphicx}

\hypersetup{pdfpagetransition={Split}}

\newcommand{\evenhead}{Author \ name}
\newcommand{\oddhead}{Article \ name}
\newcommand{\theArticleName}{Article \ name}

\newcommand{\FirstPageHeading}[1]{\thispagestyle{empty}%
\noindent\raisebox{0pt}[0pt][0pt]{\makebox[\textwidth]{\protect\footnotesize \sf }}\par}

\newcommand{\ArticleName}[1]{\renewcommand{\theArticleName}{#1}\vspace{-2mm}\par\noindent {\LARGE\bf  #1\par}}
\newcommand{\Author}[1]{\vspace{5mm}\par\noindent {\Large  #1\par} \par\vspace{2mm}\par}
\newcommand{\Address}[1]{\vspace{2mm}\par\noindent {\it #1} \par}
\newcommand{\Email}[1]{\ifthenelse{\equal{#1}{}}{}{\par\noindent {\rm E-mail: }{\it  #1} \par}}
\newcommand{\URLaddress}[1]{\ifthenelse{\equal{#1}{}}{}{\par\noindent {\rm URL: }{\tt  #1} \par}}
\newcommand{\EmailD}[1]{\ifthenelse{\equal{#1}{}}{}{\par\noindent {$\phantom{\dag}$~\rm E-mail: }{\it  #1} \par}}
\newcommand{\URLaddressD}[1]{\ifthenelse{\equal{#1}{}}{}{\par\noindent {$\phantom{\dag}$~\rm URL: }{\tt  #1} \par}}

\newcommand{\Abstract}[1]{\vspace{6mm}\par\noindent\hspace*{10mm}
\parbox{140mm}{\small {\bf Abstract.} #1}\par}
\newcommand{\Keywords}[1]{\vspace{3mm}\par\noindent\hspace*{10mm}
\parbox{140mm}{\small {\bf Key words:} \rm #1}\par}
\newcommand{\Classification}[1]{\vspace{3mm}\par\noindent\hspace*{10mm}
\parbox{140mm}{\small {\it 2020 Mathematics Subject Classification:} \rm #1}\vspace{3mm}\par}
\newcommand{\ShortArticleName}[1]{\renewcommand{\oddhead}{#1}}
\newcommand{\AuthorNameForHeading}[1]{\renewcommand{\evenhead}{#1}}

\long\def\@makecaption#1#2{
  \sbox\@tempboxa{\small \textbf{#1.}\ \ #2}%
  \ifdim \wd\@tempboxa >\hsize
    {\small \textbf{#1.}\ \ #2}\par \else
    \global \@minipagefalse
    \hb@xt@\hsize{\hfil\box\@tempboxa\hfil}%
  \fi \vskip\belowcaptionskip}

\def\numberwithin#1#2{\@ifundefined{c@#1}{\@nocounterr{#1}}{%
  \@ifundefined{c@#2}{\@nocnterr{#2}}{%
  \@addtoreset{#1}{#2}%
  \toks@\@xp\@xp\@xp{\csname the#1\endcsname}%
  \@xp\xdef\csname the#1\endcsname
    {\@xp\@nx\csname the#2\endcsname.\the\toks@}}}}
\def\E^#1{{\buildrel #1 \over\vee}}
\newtheorem{theorem}{Theorem}

{\theoremstyle{definition}

}

\usepackage{tikz}

\begin{document}

\FirstPageHeading{V.I. Gerasimenko}

\ShortArticleName{Methods for deriving kinetic equations}

\AuthorNameForHeading{V.I. Gerasimenko}

\ArticleName{\textcolor{blue!50!black}{Methods for Deriving Nonlinear Kinetic Equations\\
                                       from the Dynamics of Quantum Systems}
}

\Author{V.I. Gerasimenko$^\ast$\footnote{E-mail: \emph{gerasym@imath.kiev.ua}}}

\Address{$^\ast$ Institute of Mathematics of the NAS of Ukra\"{\i}ne,\\
    \hspace*{4mm}3, Tereshchenkivs'ka Str.,\\
    \hspace*{4mm}01601, Ky\"{\i}v-4, Ukra\"{\i}ne}

\bigskip

\Abstract{The article outlines key topics related to deriving kinetic equations and, in general,
to the relationship of the evolution equations governing quantum many-particle systems.
The main issues being considered are: the mathematical origin of quantum kinetic equations, novel
methods for deriving kinetic equations from hierarchies of evolution equations for quantum systems,
specifically focusing on kinetics within the framework of the evolution of observables, and on the
generalization of the quantum kinetic equation specified by non-separable initial states.
}

\bigskip

\Classification{35C10, 35Q20, 82C05, 82D05, 70F45.}

\makeatletter
\renewcommand{\@evenhead}{
\hspace*{-3pt}\raisebox{-7pt}[\headheight][0pt]{\vbox{\hbox to \textwidth {\thepage \hfil \evenhead}\vskip4pt \hrule}}}
\renewcommand{\@oddhead}{
\hspace*{-3pt}\raisebox{-7pt}[\headheight][0pt]{\vbox{\hbox to \textwidth {\oddhead \hfil \thepage}\vskip4pt\hrule}}}
\renewcommand{\@evenfoot}{}
\renewcommand{\@oddfoot}{}
\makeatother

\newpage
\vphantom{math}

\protect\textcolor{blue!50!black}{\tableofcontents}

\vspace{0.7cm}

\textcolor{blue!50!black}{\section{Introduction}}

One of the principal challenges in kinetic theory is understanding how to describe the evolution
of a system composed of many particles through the behavior of a typical particle. The conventional
philosophy of describing kinetic quantum evolution states that if a separable initial state is specified
by a one-particle (reduced) density operator, then at an arbitrary time, the evolution of this state in
the appropriate scaling limit can be effectively described by a one-particle density operator governed
by a nonlinear kinetic equation. Recently, there has been a resurgence of interest in deriving kinetic
equations from the dynamics of many quantum particles. One key challenge was largely due to unresolved
issues in the quantum theory of information transfer processes \cite{MW}, which are linked to the
description of the evolution of correlations within complex systems \cite{SMLR24,GG22,BGSRS22}.

The problem of deriving kinetic equations has deeper mathematical implications. These lie in the
relationships between the fundamental evolution equations that describe many-particle systems and the
heuristically formulated equations that explain the evolution of systems at various scales. The origins
of this problem can be traced back to H. Poincar\'{e} in 1905 and D. Hilbert in 1912, when it was framed
in the context of deriving hydrodynamic equations from the Boltzmann kinetic equation. This concept was
later developed by M. Bogolyubov in 1945, who derived kinetic equations from the hierarchy of fundamental
evolution equations known as the BBGKY (Bogolyubov--Born--Green--Kirkwood--Yvon) hierarchy, which describes
the evolution of states in many-particle systems. A rigorous theory for describing equilibrium states was
developed in the 1960s, while the development of a theory of nonequilibrium states began in the 1980s. We
refer to review \cite{GG23} for a synopsis of the chronology of kinetic theory. The mathematical theory
of the quantum BBGKY hierarchy originates from D. Petrina's article \cite{Pe71}. At the end of the XXth
century, the results of the development of the mathematical theory of the BBGKY hierarchy of evolution
equations and its relationship to kinetic equations for classical systems were summed up in books
\cite{CGP97,CIP,Sp}.

The approaches to deriving kinetic equations rely on our existing capabilities to construct solutions
for the BBGKY hierarchy \cite{CGP97},\cite{YHT,P2021}. Until now, such solutions and their scaling
asymptotics have been constructed using perturbation theory methods. The greatest progress has been
associated with the derivation of the Boltzmann equation for the collision dynamics of hard spheres
\cite{GG23},\cite{HP23,Ann23,GG21,GG}. It should be noted that the problem of deriving the Boltzmann
equation is, in a sense, an evergreen issue. This year, a recent contribution to this problem, the
rigorous solution of which was first published in the paper by D. Petrina and the author of this article
\cite{PG90}, was awarded a Fields Medal \cite{DHM}. The last two decades have also seen significant
progress in deriving quantum kinetic equations \cite{N28,UU33,BG47} in scaling limits of the perturbative
solution of the BBGKY hierarchy \cite{G13,G20,BPS}, in particular, the quantum Boltzmann equation in the
weak coupling limit \cite{P04,BCEP3,CH}, the nonlinear Schr\"{o}dinger equation and the Gross--Pitaevskii
equation in the mean field limit \cite{PP09,Go,EShY10,Sc,HST}.

This article formulates methods for deriving kinetic equations based on non-perturbative solutions of
hierarchies of evolution equations for quantum systems \cite{GG23d}. One of the achievements of these
methods is the construction of kinetic equations for non-separable initial states, i.e., in the presence
of correlations between initial states, as well as the description of the propagation processes of
correlations that are described by their solution \cite{GG22,BGSRS22},\cite{GSh08,GP11}. The hope is
that such approaches will eventually yield a mathematically rigorous derivation of kinetic equations
that aligns with physical intuition.

The article is structured around the key themes related to the challenge of the relationship of the
evolution equations that describe many-particle systems.
The main issues being considered are:

Hierarchies of fundamental evolution equations of quantum systems and non-perturbative methods for
constructing their solutions (Section 2);

Mathematical origins of quantum kinetic equations and the evolution of correlations governed by kinetic
equations (Section 3);

Novel methods for deriving kinetic equations from hierarchies of evolution equations of quantum systems,
specifically focusing on kinetics within the framework of the evolution of observables and the generalized
quantum kinetic equation with initial correlations (Section 4).

In Section 5, we provide a brief overview of perspectives for future research.


%
\textcolor{blue!50!black}{\section{Hierarchies of evolution equations for quantum systems}}

Many-particle systems are described in terms of concepts of observables and the state.
The mean-value functional of observables defines the duality between observables and a state.
As a consequence, there are two equivalent approaches to describing the evolution of systems:
either as the evolution of observables or as the evolution of a state.

In the paper, we will explore quantum systems of a non-fixed number of identical spinless particles,
i.e., an arbitrary but finite average number of particles that obey the Maxwell--Boltzmann statistics,
and use a system of units where $h={2\pi\hbar}=1$ is Planck's constant and $m =1$ is the particle mass.
The results can be extended to systems of many bosons or fermions, following \cite{GP11}.

We first describe the traditional approach to describing observables and the state of quantum many-particle
systems and introduce the notation needed for the subsequent presentation in terms of the nonequilibrium
grand canonical ensemble.

Let $\mathcal{H}^{\otimes n}$ be an $n$-particle Hilbert space. The Fock space over the space $\mathcal{H}$
is denoted by $\mathcal{F}_{\mathcal{H}}={\bigoplus\limits}_{n=0}^{\infty}\mathcal{H}^{\otimes n}$.
Let $\mathfrak{L}(\mathcal{H}^{\otimes n})$ be the space of bounded self-adjoint operators
$b_{n}\in\mathfrak{L}(\mathcal{H}^{\otimes n})$ with operator norm $\|.\|_{\mathfrak{L}(\mathcal{H}^{\otimes n})}$.
The self-adjoint operator $b_{n}$ is defined on the space $\mathcal{H}^{\otimes n}$,
hereinafter also denoted by the symbol $b_{n}(1,\ldots,n)$, which for arbitrary values of the indices
$(i_{1},\ldots,i_{n})\in(1,\ldots,n)$ satisfies the symmetry condition: $b_{n}(1,\ldots,n)=b_{n}(i_{1},\ldots,i_{n})$.
Let us introduce the space $\mathfrak{L}_\gamma(\mathcal{F}_{\mathcal{H}})$ of sequences of bounded operators
$b=(b_0,b_1(1),\ldots,b_n(1,\ldots,n),\ldots)$ with norm
\begin{eqnarray*}
   &&\|b\|_{\mathfrak{L}_\gamma(\mathcal{F}_{\mathcal{H}})}=
      \max_{n\geq0}\frac{\gamma^n}{n!}\,\|b_{n}\|_{\mathfrak{L}(\mathcal{H}^{\otimes n})},
\end{eqnarray*}
where $0<\gamma<1$ is a real number. The subspace of finite sequences of degenerate operators
whose kernels are infinitely differentiable functions with compact supports is denoted by
$\mathfrak{L}_0(\mathcal{F}_{\mathcal{H}})$.

Accordingly, let $\mathfrak{L}^{1}(\mathcal{H}^{\otimes n})$ be the space of trace class operators
$f_{n}\equiv f_{n}(1,\ldots,n)\in\mathfrak{L}^{1}(\mathcal{H}^{\otimes n})$, which for arbitrary
values of the indices $(i_{1},\ldots,i_{n})\in(1,\ldots,n)$ satisfy the symmetry condition:
$f_{n}(1,\ldots,n)=f_{n}(i_{1},\ldots,i_{n})$, with norm:
$\|f_{n}\|_{\mathfrak{L}^{1}(\mathcal{H}^{\otimes n})}=\mathrm{Tr}_{1,\ldots,n}|f_{n}(1,\ldots,n)|,$
where the symbol $\mathrm{Tr}_{1,\ldots,n}$ denote the partial traces of the operator $f_{n}$.
Let us introduce the space $\mathfrak{L}^{1}_\alpha(\mathcal{F}_{\mathcal{H}})$ of sequences
of trace class operators $f=(f_0,f_1(1),\ldots,f_n(1,\ldots,n),\ldots)$ with norm
\begin{eqnarray*}
   &&\|f\|_{\mathfrak{L}^{1}_\alpha(\mathcal{F}_{\mathcal{H}})}=
      \sum_{n=0}^\infty \alpha^n \mathrm{Tr}_{1,\ldots,n}|f_{n}(1,\ldots,n)|,
\end{eqnarray*}
where $\alpha>1$ is a real number. The subspace of finite sequences of degenerate operators
whose kernels are infinitely differentiable functions with compact supports will be denoted
by $\mathfrak{L}^{1}_0(\mathcal{F}_{\mathcal{H}})$.

\textcolor{blue!50!black}{\subsection{Evolution of observables of a finite number of quantum particles}}

The evolution of observables of a non-fixed but finite average number of quantum particles is described
by sequences of self-adjoint operators defined on the Fock space
$\mathcal{F}_{\mathcal{H}}={\bigoplus\limits}_{n=0}^{\infty}\mathcal{H}^{\otimes n}$
\begin{eqnarray}\label{Hes}
  &&\hskip-7mm O_n(t,1,\ldots,n)=\mathcal{G}_n(t)O_n^0(1,\ldots,n),\quad n\geq1,
\end{eqnarray}
which, for an $n$-particle quantum subsystem, are determined by the following one-parameter mapping defined
on the space $\mathfrak{L}(\mathcal{H}^{\otimes n})$:
\begin{eqnarray}\label{group}
  &&\hskip-7mm \mathcal{G}_n(t)b_n\doteq e^{itH_{n}}b_n e^{-itH_{n}},
\end{eqnarray}
where the self-adjoint operator $H_n$ is the Hamiltonian of $n$ quantum particles. This operator has the
form $H_n=\sum_{j=1}^n K(j)+\sum_{1\le j_1<j_2\le n}\Phi(j_1,j_2)$, where $K(j)$ is the kinetic energy
operator for the $j$-th particle, and $\Phi$ is the pair-interaction potential, which satisfies the Kato
conditions \cite{DL}. From now on, we take the interaction potential to be bounded.

In the space $\mathfrak{L}(\mathcal{H}^{\otimes n})$ the one-parameter mapping \eqref{group} forms a $\ast$-weakly
continuous group of operators, the infinitesimal generator of which, on the domain of its definition, coincides with
the Heisenberg operator \cite{DL}
\begin{eqnarray}\label{Heg}
  &&\hskip-7mm \mathcal{N}_{n}=
     \sum_{j=1}^{n}\mathcal{N}(j)+\sum_{j_1<j_{2}=1}^{n}\mathcal{N}_{\mathrm{int}}(j_{1},j_{2}),
\end{eqnarray}
where, for $b_n\in\mathfrak{L}_0(\mathcal{H}^{\otimes n})$, we introduced the notations:
\begin{eqnarray*}
   &&\hskip-7mm \mathcal{N}(j)b_{n}\doteq-i(b_{n}K(j)-K(j)b_{n}),\\
   &&\hskip-7mm \mathcal{N}_{\mathrm{int}}(j_{1},j_{2})b_{n}\doteq
     -i(b_{n}\Phi(j_{1},j_{2})-\Phi(j_{1},j_{2})b_{n}).
\end{eqnarray*}

The sequence of operators \eqref{Hes} is determined by the Cauchy problem for the sequence
of Heisenberg equations:
\begin{eqnarray}\label{He}
   &&\hskip-7mm \frac{\partial}{\partial t}O_n(t)=\mathcal{N}_n O_n(t), \\
   \label{Hei}
   &&\hskip-7mm O_n(t)|_{t=0}=O_n^0,\quad n\geq1,
\end{eqnarray}
where the generator of these evolution equations is given by formula \eqref{Heg}.

In view of the properties of operator groups \eqref{group}, an existence theorem holds \cite{DL}.

\medskip
\begin{theorem}
For initial data $O(0)\in \mathfrak{L}_0(\mathcal{F}_{\mathcal{H}})$, the sequence \eqref{Hes}
is the unique global in time classical solution to the Cauchy problem for the Heisenberg equations
\eqref{He},\eqref{Hei}. For arbitrary initial data in the space
$\mathfrak{L}_\gamma(\mathcal{F}_{\mathcal{H}})$, it is the unique global in time generalized solution.
\end{theorem}

A similar result also holds for sequences of unbounded operators \cite{DL}.

\textcolor{blue!50!black}{\subsection{Evolution of a state of a finite number of quantum particles}}

The mean value (mathematical expectation) at a time $t\in\mathbb{R}$ of an observable
$O(t)=(O_0,O_1(t,1),\ldots,$  $O_n(t,1,\ldots,n),\ldots)$ for quantum systems with a non-fixed number
of identical particles is defined by a continuous linear functional, which can be expressed in the
form of the following series expansion:\cite{G13}:
\begin{eqnarray}\label{ev}
   &&\hskip-8mm\langle O\rangle(t)\doteq
      (I,D(0))^{-1}\sum\limits_{n=0}^{\infty}\frac{1}{n!}\,\mathrm{Tr}_{1,\ldots,n}O_n(t)D_{n}^0\equiv\\
   &&(I,D(0))^{-1}(O(t),D(0)),\nonumber
\end{eqnarray}
where the abbreviated notation was introduced. The sequence $D(0)=(I,D_{1}^0,\ldots,D_{n}^0,\ldots)$ of
positive self-adjoint operators $D_{n}^0\in\mathfrak{L}^1(\mathcal{H}^{\otimes n})$, whose kernels are
known as density matrices, is the sequence of density operators that describe all possible states of the
quantum system with a non-fixed number of particles at the initial instant. The coefficient
$(I,D(0))={\sum\limits}_{n=0}^{\infty}\frac{1}{n!}\mathrm{Tr}_{1,\ldots,n}D_{n}^0$ is the normalizing
factor known as the grand canonical partition function.

The mean-value functional \eqref{ev} defines a duality between observables and the state.
For $D_{n}^0\in\mathfrak{L}^{1}(\mathcal{H}^{\otimes n})$ and $O_{n}(t)\in\mathfrak{L}(\mathcal{H}^{\otimes n})$,
it exists.

The sequence of density operators at instant $t\in\mathbb{R}$ we denote by $D(t)=(I,D_{1}(t),\ldots,D_{n}(t),\ldots)$.
Given the validity of the fundamental relation for representations of the mean-value functional for the observables
\begin{eqnarray}\label{evD}
   &&\hskip-7mm\langle O\rangle(t)=(I,D(0))^{-1}(O(t),D(0))=
      (I,D(0))^{-1}\sum\limits_{n=0}^{\infty}\frac{1}{n!}\,\mathrm{Tr}_{1,\ldots,n}\,\mathcal{G}_n(t)O^0_n \,D_{n}^0=\\
   &&(I,D(t))^{-1}\sum\limits_{n=0}^{\infty}\frac{1}{n!}\,\mathrm{Tr}_{1,\ldots,n}O^0_n\,\mathcal{G}^{\ast}_n(t)D_{n}^0=
      (I,D(t))^{-1}(O(0),D(t)),\nonumber
\end{eqnarray}
the density operator of an $n$-particle quantum system
\begin{eqnarray}\label{vNes}
  &&\hskip-7mm D_n(t,1,\ldots,n)=\mathcal{G}^{\ast}_n(t)D_n^0(1,\ldots,n),\quad n\geq1,
\end{eqnarray}
is determined by the group of operators adjoint to operators \eqref{group} for $f_n\in\mathfrak{L}^{1}(\mathcal{H}^{\otimes n})$
\begin{eqnarray}\label{dgroup}
   &&\hskip-7mm \mathcal{G}_n^{\ast}(t)f_n=e^{-itH_{n}}f_n e^{itH_{n}}.
\end{eqnarray}

In the space $\mathfrak{L}^1(\mathcal{H}^{\otimes n})$, the dual one-parameter mapping \eqref{dgroup}
in the sense of functional \eqref{evD} forms a strongly continuous isometric group of operators that
preserves positivity and self-adjointness of operators. The infinitesimal generator of the operator
group \eqref{dgroup} on the subspace $\mathfrak{L}^{1}_0(\mathcal{H}^{\otimes n})$ coincides with
the following von Neumann operator \cite{G13}:
\begin{eqnarray}\label{vNeg}
  &&\hskip-7mm\mathcal{N}_{n}^{\ast}=\sum_{j=1}^{n}\mathcal{N}^{\ast}(j)+
     \sum_{j_1<j_{2}=1}^{n}\mathcal{N}^{\ast}_{\mathrm{int}}(j_{1},j_{2}).
\end{eqnarray}
For $f_n\in\mathfrak{L}^1_0(\mathcal{H}^{\otimes n})$, we introduced the notations:
\begin{eqnarray*}
   &&\hskip-7mm \mathcal{N}^{\ast}(j)f_n\doteq-i(K(j)f_{n}-f_{n}K(j)),\\
   &&\hskip-7mm \mathcal{N}^{\ast}_{\mathrm{int}}(j_1,j_{2})f_n\doteq-i(\Phi(j_{1},j_{2})f_{n}-f_{n}\Phi(j_{1},j_{2})),
\end{eqnarray*}
where the operator $K(j)$ is the kinetic energy of the $j$-th particle, and $\Phi$
is the pair-interaction potential.

For further discussion, we introduce the key concept of cumulants (semi-invariants) of operator
groups \cite{GRS04}, that is, the connected part of operator groups.

The cumulant of the $n$th-order of operator groups \eqref{dgroup} is defined by the following
expansion \cite{G13}:
\begin{eqnarray}\label{dcum}
   &&\hskip-7mm \mathfrak{A}_{n}(t,1,\ldots,n)=\sum\limits_{\mathrm{P}:\,(1,\ldots,n)=
       \bigcup_i X_i}(-1)^{|\mathrm{P}|-1}({|\mathrm{P}|-1})!
      \prod\limits_{X_i\subset\mathrm{P}}\mathcal{G}^{\ast}_{|X_i|}(t,X_i),
\end{eqnarray}
where the symbol ${\sum\limits}_{\mathrm{P}:(1,\ldots,n)=\bigcup_{i} X_{i}}$ denotes the sum over
all partitions $\mathrm{P}$ of the index set $(1,\ldots,n)$ into $|\mathrm{P}|$ non-empty subsets
$X_i\subset (1,\ldots,n)$ that are mutually disjoint.
The simplest examples of cumulants of operator groups \eqref{dcum} are presented by expansions:
\begin{eqnarray*}
   &&\hskip-7mm \mathfrak{A}_{1}(t,1)=\mathcal{G}_{1}^{\ast}(t,1),\\
   &&\hskip-7mm \mathfrak{A}_{2}(t,1,2)=\mathcal{G}_{2}^{\ast}(t,1,2)-\mathcal{G}_{1}^{\ast}(t,1)\mathcal{G}^{\ast}_{1}(t,2),\\
   &&\hskip-7mm \mathfrak{A}_{3}(t,1,2,3)=\mathcal{G}_{3}^{\ast}(t,1,2,3)-\mathcal{G}_{1}^{\ast}(t,1)\mathcal{G}_{2}^{\ast}(t,2,3)-
    \mathcal{G}_{1}^{\ast}(t,2)\mathcal{G}^{\ast}_{2}(t,1,3)-\\
    &&\hskip+7mm \mathcal{G}^{\ast}_{1}(t,3)\mathcal{G}^{\ast}_{2}(t,1,2)+
    2!\,\mathcal{G}^{\ast}_{1}(t,1)\mathcal{G}^{\ast}_{1}(t,2)\mathcal{G}^{\ast}_{1}(t,3),\\
   && \vdots
\end{eqnarray*}

Cumulants of operator groups \eqref{dcum} are solutions of recursive equations known as cluster expansions
of operator groups \eqref{dgroup}
\begin{equation}\label{exp}
    \mathcal{G}^{\ast}_{n}(t,1,\ldots,n)=\sum\limits_{\mathrm{P}:\,(1,\ldots,n)=
       \bigcup_i X_i}\prod_{X_i\subset \mathrm{P}}\mathfrak{A}_{|X_i|}(t,X_i),\quad n\geq 1,
\end{equation}
where the notation of the formula \eqref{dcum} is used. It should be noted that the uniqueness of the structure
of cluster expansions is determined by the symmetry of the infinitesimal generator \eqref{vNeg} of the operator
groups \eqref{dgroup}.

For further consideration, we will also generalize the concepts of cluster and cumulant expansions of operator
groups for systems consisting of clusters of quantum particles.
We will use such abridged notations.
The symbol $(\{i_1,\ldots,i_s\})$ denotes a connected subset of indices $(i_1,\ldots,i_s)$ of the set of indices
$(1,\ldots,n)$, i.e., a set consisting of a single index element. Then the set
$(\{X_{1}\},\ldots,\{X_{|\mathrm{P}|}\})\in(1,\ldots,n)$ is the set of the $|\mathrm{P}|$ elements that are
mutually disjoint subsets $X_{i}\subset (1,\ldots,n)$ of the partition $\mathrm{P}$ of the set of indices
$(1,\ldots,n)={\bigcup\limits}_{i=1}^{|\mathrm{P}|}X_i.$

The cumulant of the $|\mathrm{P}|$th-order of operator groups \eqref{dgroup} for $|\mathrm{P}|$ clusters of
$n$ quantum particles is denoted by $\mathfrak{A}_{|\mathrm{P}|}(t,\{X_{1}\},\ldots,\{X_{|\mathrm{P}|}\})$,
and it is defined by the following expansion:
\begin{eqnarray}\label{cdcum}
   &&\hskip-7mm \mathfrak{A}_{|\mathrm{P}|}(t,\{X_1\},\ldots,\{X_{|\mathrm{P}|}\})\doteq\\
   &&\sum\limits_{\mathrm{P}^{'}:\,(\{X_1\},\ldots,\{X_{|\mathrm{P}|}\})=
      \bigcup_k Y_k}(-1)^{|\mathrm{P}^{'}|-1}({|\mathrm{P}^{'}|-1})!
      \prod\limits_{Y_k\subset\mathrm{P}^{'}}\mathcal{G}^{\ast}_{|\theta(Y_{k})|}(t,\theta(Y_{k})),\nonumber
\end{eqnarray}
where the notations are similar to that in formula \eqref{dcum} and the operator $\theta$ is the declusterization
mapping defined by $\theta(\{X_1\},\ldots,\{X_{|\mathrm{P}|}\})\doteq(1,\ldots,n)$.

Note that the cumulant \eqref{cdcum} of the $|\mathrm{P}|$th-order of operator groups for the $|\mathrm{P}|$
clusters of $n$ quantum particles is expressed in terms of the cumulants \eqref{dcum} of operator groups for
$n$ particles. A typical example of the cumulant of a cluster of $s$ particles and $n$ particles is determined
by the expansions:
\begin{eqnarray}\label{c-p}
    &&\hskip-7mm \mathfrak{A}_{1+n}(t,\{1,\ldots,s\},s+1,\ldots,s+n)=
       \sum\limits_{\scriptsize\mathrm{P}:(\{1,\ldots,s\},s+1,\ldots,s+n)=\bigcup_i X_i}
       (-1)^{|\mathrm{P}|-1}(|\mathrm{P}|-1)!\times \\
    &&\prod_{X_i\subset\mathrm{P}}\,\sum\limits_{\mathrm{P'}:\,\theta(X_{i})=\bigcup_{j_i} Y_{j_i}}\,
      \prod_{Y_{j_i}\subset\mathrm{P'}}\mathfrak{A}_{|Y_{j_i}|}(t,Y_{j_i}), \quad  n\geq0,\nonumber
\end{eqnarray}
for example,
\begin{eqnarray*}
    &&\hskip-7mm \mathfrak{A}_{1+0}(t,\{1,2\})=\mathfrak{A}_{2}(t,1,2)+
      \mathfrak{A}_{1}(t,1)\mathfrak{A}_{1}(t,2),\\
    &&\hskip-7mm \mathfrak{A}_{1+1}(t,\{1,2\},3)=\mathfrak{A}_{3}(t,1,2,3)+
      \mathfrak{A}_{2}(t,1,3)\mathfrak{A}_{1}(t,2)+
      \mathfrak{A}_{2}(t,2,3)\mathfrak{A}_{1}(t,1),\\
   && \vdots
\end{eqnarray*}

If $f_{n}\in\mathfrak{L}^1(\mathcal{H}^{\otimes n})$, then for the cumulant of $n$th-order \eqref{dcum},
the following estimate is valid
\begin{eqnarray}\label{est}
   &&\hskip-7mm\big\|\mathfrak{A}_{n}(t,1,\ldots,n)f_{n}\big\|_{\mathfrak{L}^1(\mathcal{H}^{\otimes n})}
      \leq\sum\limits_{\scriptsize\mathrm{P}:\,(1,\ldots,n)={\bigcup}_i X_i}
      (|\mathrm{P}|-1)!\,\big\|f_{n}\big\|_{\mathfrak{L}^1(\mathcal{H}^{\otimes n})}\leq \\
   &&\sum\limits_{k=1}^{n}\mathrm{s}(n,k)(k-1)!\big\|f_{n}\big\|_{\mathfrak{L}^1(\mathcal{H}^{\otimes n})}
      \leq n!\,e^{n}\,\big\|f_{n}\big\|_{\mathfrak{L}^1(\mathcal{H}^{\otimes n})},\nonumber
\end{eqnarray}
where $\mathrm{s}(n,k)$ are the Stirling numbers of the second kind.

\textcolor{blue!50!black}{\subsubsection{Evolution of states in terms of density operators}}

The evolution of a state of non-fixed, but finite average number of quantum particles, i.e.,
the sequence of density operators \eqref{vNes}, is determined by the Cauchy problem for the
sequence of the von Neumann equations:
\begin{eqnarray}\label{vNe}
   &&\hskip-7mm \frac{\partial}{\partial t}D_n(t)=\mathcal{N}^{\ast}_n D_n(t), \\
   \label{vNei}
   &&\hskip-7mm D_n(t)|_{t=0}=D_n^0,\quad n\geq 1,
\end{eqnarray}
where the generator of these evolution equations is given by formula \eqref{vNeg}.

As a consequence of the properties of operator groups \eqref{dgroup}, an existence theorem holds \cite{G13}.

\medskip
\begin{theorem}
For initial data $D(0)\in \mathfrak{L}^1_0(\mathcal{F}_{\mathcal{H}})$, the sequence \eqref{vNes} is the
unique global-in-time strong solution to the Cauchy problem for the von Neumann equations \eqref{vNe},\eqref{vNei},
whereas for arbitrary initial data from the space $\mathfrak{L}^{1}_\alpha(\mathcal{F}_{\mathcal{H}})$, it
is a unique global-in-time weak solution.
\end{theorem}

This statement also holds for interaction potentials that satisfy the Kato conditions \cite{DL}.

\textcolor{blue!50!black}{\subsubsection{Evolution of states in terms of correlation operators}}

An alternative way to describe the states of quantum particle systems involves using operators defined
as cumulant expansions of density operators, which are interpreted as correlations of particle states.

We introduce sequence $g(t)=(I,g_{1}(t,1),\ldots,g_{s}(t,1,\ldots,s),\ldots)$ of correlation operators
by means of cluster expansions of the density operators $D(t)=(I,D_{1}(t,1),\ldots,D_{n}(t,1,\ldots,n),\ldots)$:
\begin{eqnarray}\label{Dg}
    &&\hskip-7mm D_{n}(t,1,\ldots,n)=\sum\limits_{\mathrm{P}:(1,\ldots,n)=\bigcup_{i} X_{i}}
        \,\otimes_{X_i\subset \mathrm{P}}g_{|X_i|}(t,X_i),\quad n\geq1,
\end{eqnarray}
where the symbol ${\sum\limits}_{\mathrm{P}:(1,\ldots,n)=\bigcup_{i} X_{i}}$ denotes the sum over all
partitions $\mathrm{P}$ of the index set $(1,\ldots,n)$ into $|\mathrm{P}|$ non-empty subsets
$X_i\subset (1,\ldots,n)$ that are mutually disjoint.

The solutions of recurrent relations \eqref{Dg} are determined by the following expansions:
\begin{eqnarray}\label{gD}
   &&\hskip-7mm g_{n}(t,1,\ldots,n)=\sum\limits_{\mathrm{P}:(1,\ldots,n)=\bigcup_{i} X_{i}}
        (-1)^{|\mathrm{P}|-1}(|\mathrm{P}|-1)!\,\otimes_{X_i\subset \mathrm{P}}D_{|X_i|}(t,X_i), \quad n\geq1.
\end{eqnarray}
The structure of expansions \eqref{gD} is such that correlation operators can be interpreted
as cumulants (semi-invariants) of density operators \eqref{vNes}.
The origin of the structure of cluster expansions for density operators will be argued further
in subsection 2.5. The structure of cumulant expansions for states \eqref{gD} is determined by
explicitly incorporating the normalizing factor in mean-value functional \eqref{evD}, which is
represented through reduced density operators.

We will also expand the concept of cumulant expansions for correlation operators in systems with
many quantum particles, including clusters of particles.
The correlation operator \eqref{gD} for $|\mathrm{P}|$ clusters of $n$-particle system, i.e.,
$(1,\ldots,n)={\bigcup\limits}_{\mathrm{P}}\theta(\{X_{|\mathrm{P}|}\})$, we will denote by
$g_{|\mathrm{P}|}(t,\{X_{1}\},\ldots,\{X_{|\mathrm{P}|}\})$, and it is defined by the following
expansion:
\begin{eqnarray}\label{cgD}
   &&\hskip-7mm g_{|\mathrm{P}|}(t,\{X_1\},\ldots,\{X_{|\mathrm{P}|}\})\doteq\\
   &&\sum\limits_{\mathrm{P}^{'}:\,(\{X_1\},\ldots,\{X_{|\mathrm{P}|}\})=
      \bigcup_k Y_k}(-1)^{|\mathrm{P}^{'}|-1}({|\mathrm{P}^{'}|-1})!
      \,\otimes_{Y_k\subset\mathrm{P}^{'}}D_{|\theta(Y_{k})|}(t,\theta(Y_{k})),\nonumber
\end{eqnarray}
where we used the abridged notation similar to the formula \eqref{cdcum}.
In the particular case of a cluster consisting of $n$ particles, this expansion \eqref{cgD} takes the form
\begin{eqnarray*}
  &&\hskip-7mm g_{1}(t,\{1,\ldots,n\})=D_{n}(t,1,\ldots,n).
\end{eqnarray*}
According to the cluster expansions \eqref{Dg}, we derive the relation between the correlation operator of
a cluster of particles and the correlation operators of individual particles
\begin{eqnarray*}
  &&\hskip-7mm g_{1}(t,\{1,\ldots,n\})=\sum\limits_{\mathrm{P}:\,\theta(\{1,\ldots,n\})=\bigcup_{i} X_{i}}
     \,\otimes_{X_i\subset\mathrm{P}}g_{|X_{i}|}(X_{i}).
\end{eqnarray*}

By applying the cluster expansions \eqref{exp} of operator groups \eqref{dgroup} to operator
groups in the definition \eqref{cgD} of cumulants of density operators \eqref{vNes} and considering
the definition of the correlation operators at the initial instant, we derive that the evolution of
all possible states of a quantum system of non-fixed, i.e., an arbitrary but finite number of particles,
is described by means of the sequence of the following correlation operators:
\begin{eqnarray}\label{vNhsd}
   &&\hskip-7mm g_{n}(t,1,\ldots,n)=\sum\limits_{\mathrm{P}:\,(1,\ldots,n)=\bigcup_j X_j}\,
      \prod_{X_j\subset \mathrm{P}}\mathfrak{A}_{|X_j|}(t,X_j)\,
      g^{0}_{|\mathrm{P}|}(\{X_1\},\ldots,\{X_{|\mathrm{P}|}\}),\quad n\geq1,
\end{eqnarray}
where the abridged notation, similar to that used in formula \eqref{cdcum}, is applied here.
For arbitrary initial states described in terms of individual particle correlation operators,
the evolution of correlation operators can also be represented by similar expansions, which are
equivalent to the previous expansions \eqref{vNhsd}:
\begin{eqnarray}\label{vNhs}
   &&\hskip-7mm g_{n}(t,1,\ldots,n)=\sum\limits_{\mathrm{P}:\,(1,\ldots,n)=\bigcup_j X_j}
      \mathfrak{A}_{|\mathrm{P}|}(t,\{X_1\},\ldots,\{X_{|\mathrm{P}|}\})
      \otimes_{X_i\subset\mathrm{P}}g_{|X_j|}^{0}(X_j),\quad n\geq1,
\end{eqnarray}
where generating operators are cumulants of the corresponding order of operator groups (\ref{dgroup})
for $|\mathrm{P}|$ clusters of $n$ quantum particles is defined by the expansion \eqref{cdcum}.
The simplest examples of correlation operators \eqref{vNhs} are represented by such expansions:
\begin{eqnarray*}
   &&\hskip-7mm g_{1}(t,1)=\mathfrak{A}_{1}(t,1)g_{1}^{0}(1),\\
   &&\hskip-7mm g_{2}(t,1,2)=\mathfrak{A}_{1}(t,\{1,2\})g_{2}^{0}(1,2)+
     \mathfrak{A}_{2}(t,1,2)g_{1}^{0}(1)\otimes g_{1}^{0}(2),\\
   &&\vdots
\end{eqnarray*}

In particular, for separable initial states, the sequence of initial correlation operators takes the form
$g(0)=(0,g_{1}^{0},0,\ldots,0,\ldots)$. Then the evolution of correlation operators is represented as follows:
\begin{eqnarray*}
   &&\hskip-7mm g_{n}(t,1,\ldots,n)=\mathfrak{A}_{n}(t,1,\ldots,n)\,\otimes_{j=1}^{n}g_{1}^{0}(j),
   \quad n\geq1,
\end{eqnarray*}
where the generating operator $\mathfrak{A}_{n}(t)$ is the $n$th-order cumulant of operator groups
\eqref{dgroup} determined by expansion \eqref{dcum}.

Taking into consideration the explicit form of a one-particle correlation operator \eqref{vNhs}
\begin{eqnarray*}
   &&\hskip-7mm g_{1}(t,j)=\mathfrak{A}_{1}(t,j)\,g_{1}^{0}(j)=\\
   &&\mathcal{G}_1^{\ast}(t,j)g_{1}^{0}(j)=
       e^{-\frac{i}{\hbar}tK(j)}g_{1}^{0}(j) e^{\frac{i}{\hbar}tK(j)},
\end{eqnarray*}
we derive the law of the creation of correlations (the process of creating correlations from chaos)
\begin{eqnarray}\label{cch}
   &&\hskip-7mm g_{n}(t,1,\ldots,n\mid g_{1}(t))\doteq
      \mathfrak{A}_{n}(t,1,\ldots,n)\,\otimes_{j=1}^{n}(\mathfrak{A}_{1}(t,j))^{-1}g_{1}(t,j)=\\
   && \widehat{\mathfrak{A}}_{n}(t,1,\ldots,n)\,\otimes_{j=1}^{n}g_{1}(t,j),\quad n\geq2,\nonumber
\end{eqnarray}
where the cumulant $\widehat{\mathfrak{A}}_{n}(t)$ of the groups of scattering operators was introduced
\begin{eqnarray}\label{sc}
    &&\hskip-7mm\widehat{\mathcal{G}}_{n}(t)\doteq
       \mathcal{G}_{n}^{\ast}(t,1,\ldots,n)\prod_{i=1}^{n}(\mathcal{G}_{1}^{\ast}(t,i))^{-1},\quad n\geq1.
\end{eqnarray}

Note that the non-perturbative solution \eqref{vNhs} of the Cauchy problem of the von Neumann hierarchy
\eqref{vNh},\eqref{vNhi} is reduced to an iterative (perturbative) series as a result of applying analogues
of the Duhamel equations for cumulants \eqref{dcum} of operator groups \cite{G13},\cite{GG23d}.

For $f\in\mathfrak{L}^1_\alpha(\mathcal{F}_{\mathcal{H}})$, the one-parameter family of nonlinear
mappings \eqref{vNhs}
\begin{eqnarray}\label{vNhgroup}
   &&\hskip-7mm  \mathcal{G}(t;1,\ldots,n\mid f)\doteq
       \sum\limits_{\scriptsize\mathrm{P}:\,(1,\ldots,n)={\bigcup}_i X_i}
        \mathfrak{A}^{\ast}_{|\mathrm{P}|}(t,\{X_1\},\ldots,\{X_{|\mathrm{P}|}\})
        \otimes_{X_i\subset \mathrm{P}}f_{|{X_i}|}(X_i),
\end{eqnarray}
is defined, and the following estimate holds for the correlation operators
$g(t,1,\ldots,n)=\mathcal{G}(t;1,\ldots,n\mid g(0)),\,n\geq1$:
\begin{eqnarray*}
   &&\hskip-7mm  \big\|\mathcal{G}(t;1,\ldots,n\mid g(0))\big\|_{\mathfrak{L}^{1}(\mathcal{H}^{\otimes n})}\leq n!e^{2n}c^{n},
\end{eqnarray*}
where
$c\equiv e^{3}\max(1,\max_{\mathrm{P}:\,(1,\ldots,n)=\bigcup_i X_i}\|g^0_{|X_{i}|}\|_{\mathfrak{L}^{1}(\mathcal{H}^{\otimes |X_{i}|})})$,
and we denoted $g(0)=(g_0,g_1^0,\ldots,g_n^0,\ldots)$.
In the space $f\in\mathfrak{L}^1_\alpha(\mathcal{F}_{\mathcal{H}})$, the one-parameter mapping \eqref{vNhgroup} forms
a strongly continuous group of nonlinear operators. On the subspace $\mathfrak{L}^{1}_{0}(\mathcal{F}_{\mathcal{H}})$
the infinitesimal generator of the group of nonlinear operators \eqref{vNhgroup} coincides with the following operator:
\begin{eqnarray}\label{vNhg}
        &&\hskip-7mm \mathcal{N}(1,\ldots,n\mid f)\doteq\mathcal{N}^{\ast}_{n}f_{n}(1,\ldots,n)+\\
        &&\sum\limits_{\mathrm{P}:\,(1,\ldots,n)=X_{1}\bigcup X_2}\,\sum\limits_{i_{1}\in X_{1}}
           \sum\limits_{i_{2}\in X_{2}}\mathcal{N}_{\mathrm{int}}^{\ast}(i_{1},i_{2})
           f_{|X_{1}|}(t,X_{1})\otimes f_{|X_{2}|}(t,X_{2}), \nonumber
\end{eqnarray}
where the symbol ${\sum\limits}_{\mathrm{P}:\,(1,\ldots,n)=X_{1}\bigcup X_2}$ denotes the sum over all possible
partitions $\mathrm{P}$ of the index set $(1,\ldots,n)$ into two non-empty disjoint subsets $X_1$ and $X_2$, and
the operators $\mathcal{N}^{\ast}_{n}$ and $\mathcal{N}^{\ast}_{\mathrm{int}}$ are defined by the formula \eqref{vNeg}.

The evolution of a state of non-fixed, but finite average number of quantum particles, i.e.,
the sequence of correlation operators \eqref{vNhs}, is determined by the Cauchy problem for the
sequence of the von Neumann hierarchy \cite{GSh08}:
\begin{eqnarray}\label{vNh}
   &&\hskip-7mm\frac{\partial}{\partial t}g_{n}(t,1,\ldots,n)=\mathcal{N}^{\ast}_{n}(1,\ldots,n)g_{n}(t,1,\ldots,n)+\\
   &&\sum_{\scriptsize\mathrm{P}:\,(1,\ldots,n)=X_{1}\bigcup X_2}\,\sum\limits_{i_{1}\in X_{1}}
      \sum\limits_{i_{2}\in X_{2}}\mathcal{N}_{\mathrm{int}}^{\ast}(i_{1},i_{2})
      g_{|X_{1}|}(t,X_{1})\otimes g_{|X_{2}|}(t,X_{2}), \nonumber\\
 \label{vNhi}
   &&\hskip-7mm g_{n}(t)|_{t=0}=g_{n}^{0},\quad n\geq1,
\end{eqnarray}
where the symbol ${\sum\limits}_{\mathrm{P}:\,(1,\ldots,n)=X_{1}\bigcup X_2}$ means the sum over all possible
partitions $\mathrm{P}$ of the set $(1,\ldots,n)$ into two nonempty mutually disjoint subsets $X_1$ and $X_2$,
as was accepted above.
It should be noted that the von Neumann hierarchy \eqref{vNh} for correlation operators constitutes a set of
nonlinear recurrence evolution equations. We present examples of typical equations of the hierarchy \eqref{vNh}:
\begin{eqnarray*}
    &&\hskip-7mm\frac{\partial}{\partial t}g_{1}(t,1)=\mathcal{N}^{\ast}_{1}(1)\,g_{1}(t,1),\\
    &&\hskip-7mm\frac{\partial}{\partial t}g_{2}(t,1,2)=\mathcal{N}^{\ast}_{2}(1,2)\,g_{2}(t,1,2)+
       \mathcal{N}_{\mathrm{int}}^{\ast}(1,2)\,g_{1}(t,1)\otimes g_{1}(t,2),\\
    && \vdots
\end{eqnarray*}

As a consequence of the properties of operator groups \eqref{vNhgroup}, an existence theorem holds in the
space of sequences of trace-class operators \cite{G13}.

\medskip
\begin{theorem}
For bounded interaction potentials, a non-perturbative solution to the Cauchy problem for the von Neumann
hierarchy \eqref{vNh},\eqref{vNhi} is given by the sequence of correlation operators represented by the
expansions \eqref{vNhs}. For initial data $D(0)\in \mathfrak{L}^1_0(\mathcal{F}_{\mathcal{H}})$, the sequence
of correlation operators \eqref{vNes} is the globalglobal-in-time classical solution to the Cauchy problem for
the von Neumann hierarchy, whereas for arbitrary initial data from the space
$\mathfrak{L}^{1}_\alpha(\mathcal{F}_{\mathcal{H}})$, it is a global-in-time generalized solution.
\end{theorem}

Thus, the cumulant genesis of the operators for state correlations determines a cumulant structure
of the generating operators \eqref{cdcum} for expansions \eqref{vNhs} that describe the propagation
of correlations in systems of many quantum particles.

Note that the dynamics of correlations underlie the description of the state evolution, both for
finitely and infinitely many quantum particles.

\textcolor{blue!50!black}{\subsection{On the evolution of observables and states of an infinite number of particles}}

This subsection provides the classification of hierarchies of evolution equations for quantum many-particle
systems. There are two primary approaches to describing the evolution of many-particle quantum systems. The
first approach involves observables whose evolution is governed by the hierarchy of evolution equations for
reduced observables. The second approach is based on a notion of the state whose evolution is determined by
the BBGKY hierarchy. For systems with a finite number of particles, these hierarchies of evolution equations
are equivalent to the Heisenberg equations \eqref{He} for observables and the von Neumann equations \eqref{vNe}
for density operators, respectively. The advantage of these hierarchies from a mathematical viewpoint lies
in their capacity to rigorously describe the evolution of infinitely many particles whose collective behavior
is characterized by statistical features, namely, the existence of an equilibrium state, as well as the kinetic
or hydrodynamic behavior in appropriate scaling approximations.

Representations of the mean-value functional of observables \eqref{ev} underlie the classification of possible
equivalent methods for describing observables and the state of quantum many-particle systems. To construct
new representations of the mean-value functional \eqref{ev}, it is convenient to introduce the following
operators acting on sequences of operators. In the space $\mathfrak{L}_\gamma(\mathcal{F}_{\mathcal{H}})$,
we define an operator analogous to the creation operator
\begin{eqnarray}\label{a+}
   &&\hskip-7mm(\mathfrak{a}^{+}b)_{n}(1,\ldots,n)\doteq
         \sum_{j=1}^n\,b_{n-1}((1,\ldots,n)\setminus j)\otimes\mathbbm{1}_{(j)}, \quad n\geq 1,
\end{eqnarray}
where the symbol $\mathbbm{1}_{(j)}$ denotes the identity operator in the $j$ Hilbert subspace of the space
$\mathcal{H}^{\otimes n}$. On sequences of trace-class operators from the space
$\mathfrak{L}^1_\alpha(\mathcal{F}_{\mathcal{H}})$ we define an adjoint operator to the operator \eqref{a+}
in the sense of the functional \eqref{ev}, which is an analogue to the annihilation operator,
\begin{eqnarray}\label{a}
   &&\hskip-7mm(\mathfrak{a}f)_{n}(1,\ldots,n)\doteq\mathrm{Tr}_{n+1}\,f_{n+1}(1,\ldots,n,n+1),\quad n\geq 1.
\end{eqnarray}
Then, for $b\in\mathfrak{L}_\gamma(\mathcal{F}_{\mathcal{H}})$ and
$f\in\mathfrak{L}^1_\alpha(\mathcal{F}_{\mathcal{H}})$ by virtue of the equalities:
$(b,f)=(e^{\mathfrak{a^{+}}}e^{-\mathfrak{a^{+}}}b,f)=(e^{-\mathfrak{a^{+}}}b,e^{\mathfrak{a}}f),$
for mean-value functional \eqref{ev}, we derive the following representation:
\begin{eqnarray}\label{rev}
   &&\hskip-7mm\langle O\rangle(t)=(I,D(0))^{-1}(O(t),D(0))=(B(t),F(0)),
\end{eqnarray}
where the coefficient $(I,D(0))^{-1}$ is the normalizing factor and the sequence of reduced observables
is given by the formula
\begin{eqnarray}\label{BO}
   &&\hskip-7mm B(t)=e^{-\mathfrak{a^{+}}}\mathcal{G}(t)O(0),
\end{eqnarray}
and the sequence of reduced density operators is given by the following expression
\begin{eqnarray}\label{FD}
   &&\hskip-7mm F(0)=(I,D(0))^{-1}e^{\mathfrak{a}}D(0).
\end{eqnarray}
Note that the description of observables and the system state in terms of the corresponding reduced operators
is a result of the series resummation in expression \eqref{ev}, taking into account the normalization factor
$(I,D(0))^{-1}$.

Thus, according to the definition of operator \eqref{a+}, the sequence of reduced observables \eqref{BO}
in component-wise form is given by the following expansions:
\begin{eqnarray}\label{rBO}
  &&\hskip-7mm B_s(t,1,\ldots,s)=\\
  &&\sum_{n=0}^s\,\frac{(-1)^n}{n!}\sum_{j_1\neq\ldots\neq j_{n}=1}^s
      (\mathcal{G}(t)O(0))_{s-n}((1,\ldots,s)\setminus (j_1,\ldots,
      j_{n}))\otimes\mathbbm{1}_{(j_1)}\otimes\ldots\otimes\mathbbm{1}_{(j_n)},\quad s\geq1.\nonumber
\end{eqnarray}

It should be emphasized that for the mean-value functional \eqref{rev} for the reduced observables,
a fundamental relation holds with its equivalent representation
\begin{eqnarray}\label{revF(t)}
   &&\hskip-7mm (B(t),F(0))=(B(0),F(t)),
\end{eqnarray}
where according to the definition of operators \eqref{a} and \eqref{FD}, the sequence of reduced density
operators at instant $t\in\mathbb{R}$ in component-wise form is represented by the following expansions
into series:
\begin{eqnarray}\label{rFD}
    &&\hskip-7mm F_{s}(t,1,\ldots,s)\doteq (I,D(0))^{-1}\sum\limits_{n=0}^{\infty}\frac{1}{n!}\,
         \mathrm{Tr}_{s+1,\ldots,s+n}\,(\mathcal{G}^\ast(t)D(0))_{s+n}(1,\ldots,s+n),\quad s\geq1.
\end{eqnarray}
As was noted above, observables and a state of a system with a finite number of particles are described
by one-component sequences. Consequently, the reduced observables and reduced density operators are
represented by finite sequences of operators.

The mean-value functionals \eqref{revF(t)} define a duality between reduced observables and the reduced
density operators. For $F_{n}^0\in\mathfrak{L}^{1}(\mathcal{H}^{\otimes n})$, and
$B_{n}^0\in\mathfrak{L}(\mathcal{H}^{\otimes n}),\,n\geq 1$, they exist.

\textcolor{blue!50!black}{\subsection{Hierarchy of evolution equations for reduced observables}}

As noted above, one of the approaches to describing the evolution of quantum many-particle systems is based
on sequences $B(t)=(B_0,B_{1}(t),\ldots,B_{s}(t),\ldots)\in\mathfrak{L}_{\gamma}(\mathcal{F}_\mathcal{H})$
of reduced observables \eqref{BO}. By applying the cluster expansions of operator groups \eqref{group}:
\begin{eqnarray}\label{cexp}
   &&\hskip-7mm \mathcal{G}_{s}(t,(1,\ldots,s)\setminus(j_1,\ldots,j_{n}),j_1,\ldots,j_{n})=\\
   &&\sum\limits_{\mathrm{P}:\,(\{(1,\ldots,s)\setminus(j_1,\ldots,j_{n})\},\,j_1,\ldots,j_{n})=
     \bigcup_i X_i}\,\prod\limits_{X_i\subset\mathrm{P}}\mathfrak{A}_{|X_i|}(t,X_i), \quad  n\geq 0. \nonumber
\end{eqnarray}
to the operator groups in the definition of reduced observables \eqref{rBO}, and by considering the definition
of reduced observables at the initial instant, we derive that the evolution of reduced observables for a quantum
system with a non-fixed number of particles is described by the following sequence of operators:
\begin{eqnarray}\label{sroh}
    &&\hskip-7mm B_{s}(t,1,\ldots,s)=\sum_{n=0}^s\,\frac{1}{n!}\sum_{j_1\neq\ldots\neq j_{n}=1}^s
      \mathfrak{A}_{1+n}(t,\{(1,\ldots,s)\setminus(j_1,\ldots,j_{n})\},\\
    &&j_1,\ldots,j_{n}\big)\,
      B_{s-n}^0((1,\ldots,s)\setminus(j_1,\ldots,j_{n}))\otimes\mathbbm{1}_{(j_1)}\ldots\otimes\mathbbm{1}_{(j_n)},
      \quad s\geq1,\nonumber
\end{eqnarray}
where the generating operators are the cumulants of the corresponding order for the operator groups \eqref{group}
associated with a cluster of $s-n$ quantum particles and $n$ individual particles:
\begin{eqnarray}\label{rocum}
    &&\hskip-7mm \mathfrak{A}_{1+n}(t,\{(1,\ldots,s)\setminus(j_1,\ldots,j_{n})\},j_1,\ldots,j_{n})\doteq\\
    &&\sum\limits_{\mathrm{P}:\,(\{(1,\ldots,s)\setminus(j_1,\ldots,j_{n})\},\,j_1,\ldots,j_{n})={\bigcup}_i X_i}
       (-1)^{\mathrm{|P|}-1}({\mathrm{|P|}-1})!\prod_{X_i\subset\mathrm{P}}
       \mathcal{G}_{|\theta(X_i)|}(t,\theta(X_i)), \quad n\geq 0,\nonumber
\end{eqnarray}
and the notation used here is similar to that in formula \eqref{cdcum}. We give the simplest examples of
expansions \eqref{sroh} for reduced observables:
\begin{eqnarray*}
   &&\hskip-7mm B_{1}(t,1)=\mathfrak{A}_{1}(t,1)B_{1}^0(1),\\
   &&\hskip-7mm B_{2}(t,1,2)=\mathfrak{A}_{1}(t,\{1,2\})B_{2}^0(1,2)+
         \mathfrak{A}_{2}(t,1,2)(B_{1}^0(1)\otimes\mathbbm{1}_{(2)}+\mathbbm{1}_{(1)}\otimes B_{1}^0(2)),\\
   &&\vdots
\end{eqnarray*}

It should also be noted that observables of many-particle quantum systems have a certain structure.
For example, the $k$-ary type observable has the structure
$O_n^{(k)}(1,\ldots,n)=\sum_{i_1<\ldots<i_k=1}^n o_k(i_1,\ldots,i_k),$  $n\geq 1$.
Therefore, the one-component sequences of reduced observables $B^{(1)}(0)=(0,o_{1}(1),0,\ldots)$
correspond to additive-type observables \cite{BG}. If the initial reduced observable is specified
by a sequence of additive-type operators $B^{(1)}(0)=(0,b_{1},0,\ldots)$, then the structure of
expansions \eqref{sroh} takes the form:
\begin{eqnarray*}\label{t-1}
    &&\hskip-7mm B_{s}^{(1)}(t,1,\ldots,s)=\mathfrak{A}_{s}(t,1,\ldots,s)\sum_{j=1}^s \mathbbm{1}_{(1)}\otimes\ldots
        \otimes\mathbbm{1}_{(j-1)}\otimes b_{1}(j)\otimes\mathbbm{1}_{(j+1)}\otimes\ldots
        \otimes\mathbbm{1}_{(s)}, \quad s\geq 1,
\end{eqnarray*}
whose generating operator $\mathfrak{A}_{s}(t)$ is the $s$th-order cumulant of operator groups \eqref{group}
associated with $s$ individual particles.
In the case of a sequence of initial reduced observables of $k$-ary type, $k\geq2$, operators
$B^{(k)}(0)=(0,\ldots,0,b_{k},0,\ldots)$, the structure of expansions \eqref{sroh} takes the form:
\begin{eqnarray*}\label{t-k}
   &&\hskip-7mm B_{s}^{(k)}(t)=0, \quad 1\leq s<k,\\
   &&\hskip-7mm B_{s}^{(k)}(t,1,\ldots,s)=\frac{1}{(s-k)!}\sum_{j_1\neq\ldots\neq j_{s-k}=1}^s
         \mathfrak{A}_{1+s-k}\big(t,\{(1,\ldots,s)\setminus (j_1,\ldots,j_{s-k})\},\nonumber\\
   && j_1,\ldots,j_{s-k}\big)b_{k}((1,\ldots,s)\setminus(j_1,\ldots,j_{s-k}))
\otimes\mathbbm{1}_{(j_1)}\otimes \ldots\otimes\mathbbm{1}_{(j_{s-k})},\quad s\geq k,\nonumber
\end{eqnarray*}
where the generating operator $\mathfrak{A}_{1+s-k}(t)$ is the $(1+s-k)$th-order cumulant of operator
groups \eqref{group}.

Provided that $\gamma < e^{-1}$, the following inequality holds for the sequence of reduced observables
(\ref{sroh}) in the space $\mathfrak{L}_{\gamma}(\mathcal{F}_{\mathcal{H}})$:
\begin{eqnarray*}\label{Bes}
    &&\hskip-7mm \|B(t)\|_{\mathfrak{L}_{\gamma}(\mathcal{F}_{\mathcal{H}})}\leq
          e^2\,(1-\gamma e)^{-1}\,\|B(0)\|_{\mathfrak{L}_{\gamma}(\mathcal{F}_{\mathcal{H}})}.
\end{eqnarray*}

The evolution of observables of non-fixed, but finite average number of quantum particles, i.e.,
the sequence of reduced observables \eqref{sroh}, is governed by the Cauchy problem to the following
hierarchy of evolution equations \cite{BG}:
\begin{eqnarray}\label{roh}
     &&\hskip-7mm\frac{d}{dt}B(t)=\mathcal{N}B(t)+\big[\mathcal{N},\mathfrak{a}^+\big]B(t),\\
  \label{rohi}
     &&\hskip-7mm B(t)|_{t=0}=B(0),
\end{eqnarray}
where the generator of this hierarchy of evolution equations is the Heisenberg generator \eqref{Heg}
together with the commutator of this generator and the operator \eqref{a+}, which component-wise has
the form
\begin{eqnarray*}
    &&\hskip-7mm (\big[\mathcal{N},\mathfrak{a}^+\big]B(t))_{s}(1,\ldots,s)
       =\sum_{j_1\neq j_{2}=1}^s\mathcal{N}_{\mathrm{int}}(j_1,j_{2})B_{s-1}(t,(1,\ldots,s)\setminus j_1)
       \otimes \mathbbm{1}_{(j_1)},\quad s\geq 2.
\end{eqnarray*}
In component-wise form, the hierarchy of evolution equations \eqref{roh} for reduced observables
constitutes, in fact, a system of recurrence evolution equations. Let us present examples of typical
equations from this hierarchy:
\begin{eqnarray*}
    &&\hskip-7mm\frac{\partial}{\partial t}B_{1}(t,1)=\mathcal{N}_{1}(1)\,B_{1}(t,1),\\
    &&\hskip-7mm\frac{\partial}{\partial t}B_{2}(t,1,2)=\big({\sum\limits}_{j=1}^{2}\mathcal{N}(j)+
              \mathcal{N}_{\mathrm{int}}(1,2)\big)\,B_{2}(t,1,2)+\\
    &&\mathcal{N}_{\mathrm{int}}(1,2)\,\big(B_{1}(t,1)\otimes\mathbbm{1}_{(2)}+ \mathbbm{1}_{(1)}\otimes B_{1}(t,2)\big),\\
    && \vdots
\end{eqnarray*}
Note that the structure of the generator of the hierarchy of evolution equations \eqref{roh} for quantum
particle systems with an arbitrary $n$-body interaction potential is given by:
$e^{-{\mathfrak{a}^+}}\mathcal{N}e^{{\mathfrak{a}^+}}$.
For a two-body interaction potential, the following equality holds \cite{G13}:
$e^{-{\mathfrak{a}^+}}\mathcal{N}e^{{\mathfrak{a}^+}}=\mathcal{N}+\big[\mathcal{N},\mathfrak{a}^+\big]$.

In the space $\mathfrak{L}_{\gamma}(\mathcal{F}_{\mathcal{H}})$ the following existence theorem holds \cite{GG23d}.

\medskip
\begin{theorem}
For bounded interaction potentials, a non-perturbative solution of the Cauchy problem \eqref{roh},\eqref{rohi}
is represented by expansions (\ref{sroh}), in which the generating operators are the cumulants of the operator
groups \eqref{group} of the corresponding order.
Provided that $\gamma<e^{-1}$, for initial data $B(0) \in \mathfrak{L}_{0}(\mathcal{F}_{\mathcal{H}})$, the
sequence of reduced observables \eqref{sroh} is the unique globalglobal-in-time classical solution, whereas
for arbitrary initial data $B(0)\in\mathfrak{L}_{\gamma}(\mathcal{F}_{\mathcal{H}})$, it is the unique
global-in-time generalized solution.
\end{theorem}

The proof of this statement is analogous to that of the existence theorem for the Heisenberg equations \eqref{He}
in the space of sequences of bounded operators. A similar result also holds for sequences of unbounded
operators \cite{DL}.

Thus, the cumulant background of the cluster expansions \eqref{cexp} for operator groups of a non-pertur\-bative
solution to the Heisenberg equations \eqref{He} determines a cumulant structure of the generating operators of
expansions \eqref{sroh} for the non-perturbative solution of the hierarchy of evolution equations \eqref{roh}
for reduced observables.
It should be emphasized that the uniqueness of the cluster expansion structure is determined by the symmetry
of the infinitesimal generator \eqref{Heg} of the group of operators \eqref{group}.

\textcolor{blue!50!black}{\subsection{Evolution of states of an infinite number of quantum particles}}

As noted above, one more approach to describing the evolution of quantum many-particle systems
is based on sequences of reduced density operators \eqref{FD}. By applying the cluster expansions
\eqref{exp} of operator groups \eqref{dgroup}:
\begin{eqnarray}\label{dcexp}
   &&\hskip-7mm \mathcal{G}_{s+n}^\ast(t,1,\ldots,s,s+1,\ldots,s+n)=\\
   &&\sum\limits_{\mathrm{P}:\,(\{1,\ldots,s\},\,s+1,\ldots,s+n)=
       \bigcup_i X_i}\,\prod\limits_{X_i\subset\mathrm{P}}\mathfrak{A}_{|X_i|}(t,X_i),\quad n\geq 0.\nonumber
\end{eqnarray}
to the operator groups in the definition of reduced density operators \eqref{rFD}, and by considering
the definition of reduced density operators at the initial instant, we derive that the evolution of
the reduced density operators for a quantum system with a non-fixed number of particles is described
by the following sequence of operators:
\begin{eqnarray}\label{sdh}
    &&\hskip-7mm F_s(t,1,\ldots,s)=\\
    &&\sum_{n=0}^\infty\frac{1}{n!}\,\mathrm{Tr}_{s+1,\ldots,s+n}\,
       \mathfrak{A}_{1+n}(t,\{1,\ldots,s\},s+1,\ldots,s+n)F_{s+n}^0(1,\ldots,s+n),\quad s\geq1,\nonumber
\end{eqnarray}
where the generating operators are the cumulants of the corresponding order for the operator
groups \eqref{dgroup} associated with a cluster of $s$ particles and $n$ individual particles
\begin{eqnarray}\label{drcum}
    &&\hskip-7mm \mathfrak{A}_{1+n}(t,\{1,\ldots,s\},s+1,\ldots,s+n)\doteq\\
    &&\sum\limits_{\mathrm{P}:\,(\{1,\ldots,s\},\,s+1,\ldots,s+n)={\bigcup}_i X_i}
       (-1)^{\mathrm{|P|}-1}({\mathrm{|P|}-1})!\prod_{X_i\subset\mathrm{P}}
       \mathcal{G}^\ast_{|\theta(X_i)|}(t,\theta(X_i)),\nonumber
\end{eqnarray}
and the notation used here is similar to that in formula \eqref{dcum}.

Provided that $\alpha>e$, by estimate \eqref{est}, the series \eqref{sdh} converges in the norm
of the space $\mathfrak{L}^{1}_{\alpha}(\mathcal{F}_{\mathcal{H}})$ and the following inequality holds:
\begin{eqnarray*}\label{Fes}
    &&\hskip-7mm \|F(t)\|_{\mathfrak{L}^{1}_{\alpha}(\mathcal{F}_{\mathcal{H}})}\leq
    e^{2}(1-\frac{e}{\alpha})^{-1}\|F(0)\|_{\mathfrak{L}^{1}_{\alpha}(\mathcal{F}_{\mathcal{H}})}.
\end{eqnarray*}
The parameter $\alpha$ can be interpreted as the inverse of the average number of particles.

Based on an alternative approach to describing state evolution by the von Neumann hierarchy \eqref{vNh},
the reduced density operators can also be defined through of correlation operators \eqref{vNhs}.
The definition of the reduced density operators \eqref{rFD} represented in terms of
the correlation operators \eqref{vNhs} by the following series expansion \cite{GG22}:
\begin{eqnarray}\label{Fg}
    &&\hskip-7mm F_{s}(t,1,\ldots,s)\doteq \sum\limits_{n=0}^{\infty}\frac{1}{n!}\,
       \mathrm{Tr}_{\mathrm{s+1,\ldots,s+n}}\,g_{1+n}(t,\{1,\ldots,s\},s+1,\ldots,s+n), \quad s\geq1,
\end{eqnarray}
where the correlation operators $g_{1+n}(t),\, n\geq0,$ of a cluster of $s$ particles and $n$ individual
particles are determined similar to expansions \eqref{vNhs}, namely
\begin{eqnarray}\label{vNhc-ps}
   &&\hskip-7mm g_{1+n}(t,\{1,\ldots,s\},s+1,\ldots,s+n)=\\
   && \sum\limits_{\mathrm{P}:\,(\{1,\ldots,s\},s+1,\ldots,s+n)=\bigcup_j X_j}\,
      \mathfrak{A}_{|\mathrm{P}|}(t,\{X_1\},\ldots,\{X_{|\mathrm{P}|}\})\,
      \otimes_{X_i\subset\mathrm{P}}g_{|X_j|}^{0}(X_j),\quad n\geq 0,\nonumber
\end{eqnarray}
and the notation adopted in formula \eqref{dcum} is used.
The redefinition of reduced density operators \eqref{Fg} naturally arises from their definition within the
grand canonical ensemble \eqref{rFD} when explicitly taking into account the normalization factor \cite{GG22}.

Traditionally, the evolution of many-particle systems is described within the framework of the evolution
of the state governed by the BBGKY hierarchy for reduced density operators \cite{Pe71},\cite{G13},\cite{BPS}.

The cluster expansions \eqref{exp} of operator groups \eqref{dgroup} underlie the classification of possible
solution representations \cite{GG23d} of the Cauchy problem of the BBGKY hierarchy for quantum many-particle system.
In particular, the non-perturbative solution \eqref{sdh} of the BBGKY hierarchy for many quantum particles can be
represented in the form of the perturbation (iteration) series as a result of applying analogs of the Duhamel
equation to cumulants \eqref{dcum} of operator groups \eqref{dgroup}
\begin{eqnarray*}
   &&\hskip-7mm F(t)=\sum\limits_{n=0}^{\infty}\,\int\limits_{0}^{t} dt_{1}\ldots\int\limits_{0}^{t_{n-1}}dt_{n}
        \mathcal{G}^\ast(t-t_{1})\big[\mathfrak{a},\mathcal{N}^\ast\big]\mathcal{G}^\ast(t_1-t_2)\ldots
         \mathcal{G}^\ast(t_{n-1}-t_n)\big[\mathfrak{a},\mathcal{N}^\ast\big]\mathcal{G}^\ast(t_{n})F(0).\nonumber
\end{eqnarray*}
Conventionally, perturbation theory methods are used to study the problem of the rigorous derivation of quantum
kinetic equations in the scaling limits of the solution of the BBGKY hierarchy \cite{G13},\cite{BPS}.

\textcolor{blue!50!black}{\subsubsection{The BBGKY hierarchy for reduced density operators}}

The evolution of all possible states of non-fixed but finite average number of quantum particles, i.e.,
the sequence of reduced density operators \eqref{sdh} is governed by the Cauchy problem of the BBGKY
hierarchy \cite{G13}:
\begin{eqnarray}\label{dh}
     &&\hskip-7mm\frac{d}{dt}F(t)=\mathcal{N}^\ast F(t)+\big[\mathfrak{a},\mathcal{N}^\ast\big]F(t),\\
  \label{dhi}
     &&\hskip-7mm F(t)|_{t=0}=F(0),
\end{eqnarray}
where the generator of the BBGKY hierarchy is the von Neumann generator \eqref{vNeg} together
with the commutator of this generator and the operator \eqref{a}, which component-wise has the form
\begin{eqnarray*}
   &&\hskip-7mm (\big[\mathfrak{a},\mathcal{N}^\ast \big]F(t))_{s}(1,\ldots,s)=\sum_{j=1}^s\,\mathrm{Tr}_{s+1}\,
       \mathcal{N}^\ast_{\mathrm{int}}(j,s+1)F_{s+1}(t,1,\ldots,s+1),\quad s\geq 1.
\end{eqnarray*}
Note that the structure of the generator of the BBGKY hierarchy for quantum particle systems with an arbitrary
$n$-body interaction potential is given by: $e^\mathfrak{a}\mathcal{N}^\ast e^{-\mathfrak{a}}$ \cite{G13}.
For the two-body interaction potential, the following equality holds:
$e^\mathfrak{a}\mathcal{N}^\ast e^{-\mathfrak{a}}=\mathcal{N}^\ast+[\mathfrak{a},\mathcal{N}^\ast]$.

In the space $\mathfrak{L}_{\alpha}(\mathcal{F}_{\mathcal{H}})$ the following statement holds \cite{GG23d}.

\medskip
\begin{theorem}
For bounded interaction potentials, a non-perturbative solution of the Cauchy problem to the Cauchy problem
of the BBGKY hierarchy \eqref{dh},\eqref{dhi} is determined by the sequence of reduced density operators
represented by series expansions \eqref{sdh}.
Provided that $\alpha>e$, for initial data $F(0)\in \mathfrak{L}^{1}_{0}(\mathcal{F}_{\mathcal{H}})$, the
sequence of reduced density operators \eqref{sdh} is the unique strong solution, and for arbitrary initial
data $F(0)\in \mathfrak{L}^{1}_{\alpha}(\mathcal{F}_{\mathcal{H}})$, it is a weak global-in-time solution.
\end{theorem}

It should be emphasized that the cumulant background of the cluster expansions \eqref{dcexp} for operator
groups of a non-perturbative solution to the von Neumann equations \eqref{vNe} determines a cumulant structure
for the generating operators of the series expansions \eqref{sdh} for the non-perturbative solution of the BBGKY
hierarchy \eqref{dh}. Based on an alternative approach to describing state evolution by the von Neumann hierarchy
\eqref{vNh}, the reduced density operators are determined by the correlation operators \eqref{Fg}. Then the cumulant
structure of the non-perturbative solution to the von Neumann hierarchy \eqref{vNhs} induces the cumulant structure
of the series expansion of the solution \eqref{sdh} to the Cauchy problem for the BBGKY hierarchy \eqref{dh}, which
is adopted to describe systems with an infinite number of particles; in other words, the dynamics of quantum systems
with an infinite number of particles is generated by the dynamics of correlations.

\textcolor{blue!50!black}{\subsubsection{The hierarchy of nonlinear evolution equations for reduced correlation operators}}

It is well known that, on a microscopic scale, the macroscopic characteristics of fluctuations
of observables are directly determined by the reduced correlation operators \cite{G13}.

Traditionally reduced correlation operators are introduced by the cluster expansions of the reduced
density operators:
\begin{eqnarray}\label{FG}
   &&\hskip-7mm F_{s}(t,1,\ldots,s)=\sum\limits_{\mathrm{P}:(1,\ldots,s)=\bigcup_{i}X_{i}}
      \otimes_{X_i\subset\mathrm{P}} G_{|X_i|}(t,X_i), \quad s\geq1,
\end{eqnarray}
where, as above, the symbol ${\sum\limits}_{\mathrm{P}:(1,\ldots,s)=\bigcup_{i} X_{i}}$ is the sum
over all possible partitions $\mathrm{P}$ of the set $(1,\ldots,s)$ into $|\mathrm{P}|$ nonempty
mutually disjoint subsets $X_i\subset(1,\ldots,s)$. As a consequence of this, the solutions of
recurrence relations \eqref{FG} are represented through reduced density operators as follows:
\begin{eqnarray}\label{GF}
   &&\hskip-7mm G_{s}(t,1,\ldots,s)=\sum\limits_{\mathrm{P}:(1,\ldots,s)=\bigcup_{i}X_{i}}
     (-1)^{|\mathrm{P}|-1}(|\mathrm{P}|-1)!\otimes_{X_i\subset\mathrm{P}}F_{|X_i|}(t,X_i),\quad s\geq1.
\end{eqnarray}
The structure of this expansions \eqref{GF} is such that the reduced correlation operators are cumulants
(semi-invariants) of the reduced density operators \eqref{sdh}. Consequently, these operators \eqref{GF}
are interpreted as correlations between the states of individual particles.

Assuming as a basis an alternative approach to the description of the evolution of the state within
the framework of correlation operators \eqref{vNh}, then the reduced correlation operators \eqref{GF}
are defined through of correlation operators \cite{G13}:
\begin{eqnarray}\label{Gg}
   &&\hskip-7mm G_{s}(t,1,\ldots,s)\doteq \sum\limits_{n=0}^{\infty}\frac{1}{n!}\,
      \mathrm{Tr}_{s+1,\ldots,s+n}\,g_{s+n}(t,1,\ldots,s+n),\quad s\geq1,
\end{eqnarray}
where the generating operators $g_{s+n}(t,1,\ldots,s+n),\,n\geq0,$ are determined by expansions \eqref{vNhs}
for correlation operators of $s+n$ individual particles.
In particular, for the reduced correlation operators \eqref{Gg} with a separable initial state, the following
equality holds:
\begin{eqnarray*}
   &&\hskip-7mm G_{s}^{0}(1,\ldots,s)\doteq\sum\limits_{n=0}^{\infty}\frac{1}{n!}\,
      \mathrm{Tr}_{s+1,\ldots,s+n}\,g_{1}^{0}\,\delta_{s+n,1}=g_{1}^{0}(1)\delta_{s,1},\quad s\geq 1,
\end{eqnarray*}
where $\delta{s,1}$ is a Kronecker symbol.

Considering that the evolution of correlations for a separable initial state is governed by formula
\eqref{cch}, namely
\begin{eqnarray*}
   &&\hskip-7mm G_{s}(t,1,\ldots,s)=\sum\limits_{n=0}^{\infty}\frac{1}{n!}\,\mathrm{Tr}_{s+1,\ldots,s+n}\,
      \widehat{\mathfrak{A}}_{s+n}(t,1,\ldots,s+n)\,\otimes_{j=1}^{s+n}\mathcal{G}_1^{\ast}(t,j)G_{1}^{0}(j),
\end{eqnarray*}
where $\widehat{\mathfrak{A}}_{n}(t)$ is the cumulant of the groups of scattering operators \eqref{sc},
we derive the law of creation correlations from chaos in a system with either infinitely or finitely many quantum
particles
\begin{eqnarray}\label{rcch}
   &&\hskip-7mm G_{s}(t,1,\ldots,s)=\sum\limits_{n=0}^{\infty}\frac{1}{n!}\,\mathrm{Tr}_{s+1,\ldots,s+n}\,
       \mathfrak{A}_{s+n}(t,1,\ldots,s+n)\,\otimes_{j=1}^{s+n}G_{1}^{0}(j),
\end{eqnarray}
where the generating operator $\mathfrak{A}_{s+n}(t)$ of series \eqref{rcch} is the $(s+n)$th-order cumulant
\eqref{dcum} of operator groups \eqref{dgroup}.

The evolution of all possible states of a quantum system, as an equivalent method compared with the reduced
density operators, is described by means of the sequence
$G(t)=(I,G_{1}(t),\ldots,G_{s}(t),\ldots)\in\mathfrak{L}^{1}(\mathcal{F}_\mathcal{H})$ of the reduced correlation
operators $G_{s}(t)=G_{s}(t,1,\ldots,s),\,s\geq1$, governed by the Cauchy problem of the following hierarchy of
nonlinear evolution equations \cite{G13}:
\begin{eqnarray}\label{rch}
     &&\hskip-7mm\frac{d}{dt}G(t)=\mathcal{N}(\cdot\mid G(t))+\mathfrak{a}\mathcal{N}(\cdot\mid G(t))
             -\mathcal{N}(\cdot\mid \mathfrak{a}G(t)),\\
      \label{rchi}
     &&\hskip-7mm G(t)|_{t=0}=G(0),
\end{eqnarray}
where the generator of this hierarchy of nonlinear evolution equations is the generator of the von Neuman hierarchy
\eqref{vNhg} together with the commutator of this generator and the operator \eqref{a}, which component-wise has
the form
\begin{eqnarray*}
    &&\hskip-7mm (\mathfrak{a}\mathcal{N}(\cdot\mid G(t)))_s(1,\ldots,s) -\mathcal{N}(1,\ldots,s\mid \mathfrak{a}G(t))=\\
    &&\mathrm{Tr}_{s+1}\sum_{j=1}^{s}\,\mathcal{N}^{\ast}_{\mathrm{int}}(j,s+1)\big(G_{s+1}(t,1,\ldots,s+1)+\\
    &&\sum_{\mbox{\scriptsize$\begin{array}{c}\mathrm{P}:(1,\ldots,s+1)=X_1\bigcup X_2,\\j\in
      X_1;s+1\in X_2\end{array}$}}\,G_{|X_1|}(t,X_1)\otimes G_{|X_2|}(t,X_2)\big),
\end{eqnarray*}
where the notations is similar to the formula \eqref{vNhg}, and the symbol
$\sum_{\mbox{\scriptsize$\begin{array}{c}\mathrm{P}:(1,\ldots,s+1)=X_1\bigcup X_2,\\j\in X_1;s+1\in X_2\end{array}$}}$
means the sum over all possible partitions of the set $(1,\ldots,s+1)$ into two mutually disjoint subsets $X_1$ and
$X_2$ such that $j$ particle belongs to the subset $X_1$ and $s+1$ particle belongs to $X_2$.

The statement below is true \cite{G17}.

\medskip
\begin{theorem}
For quantum many-particle systems with bounded interaction potentials and separable initial data
$G(0)=(I, G_{1}^{0},0,\ldots,0,\ldots)$, the non-perturbative solution to the Cauchy problem for
the hierarchy of nonlinear evolution equations \eqref{rch}, \eqref{rchi} is determined by a sequence
of reduced correlation operators represented by the expansions \eqref{rcch}.
Provided that $\alpha>e$, for initial data $G(0)\in \mathfrak{L}^{1}_{0}(\mathcal{F}_{\mathcal{H}})$,
the sequence of reduced correlation operators \eqref{rcch} is the unique global-in-time strong solution,
and for arbitrary initial data $G(0)\in \mathfrak{L}^{1}_{\alpha}(\mathcal{F}_{\mathcal{H}})$, it is
a weak global-in-time solution.
\end{theorem}

The outline of the proof for this assertion is analogous to the proof of the existence theorem
for the von Neumann hierarchy \eqref{vNh} in the space of sequences of trace-class operators.

\textcolor{blue!50!black}{\section{Mathematical origins of quantum kinetic equations}}

One of the challenges of kinetic theory is to describe the evolution of a system compose of many particles
within the framework of a typical particle's state, which is governed by a kinetic equation.

In circumstances when the initial state is specified by a one-particle density operator, the mean-value
functional of observables at any given moment can be represented by a one-particle density operator governed
by a non-Markovian nonlinear evolution equation. In other words, for such initial data, the Cauchy problem
for the BBGKY hierarchy is equivalent to the Cauchy problem for the nonlinear kinetic equation, and a sequence
of reduced correlation functionals of a state is determined explicitly by the solution of this kinetic equation.
The generating operators for expansions of the reduced correlation functionals of a state are determined by a
solution of the kinetic cluster expansions (analogs of equilibrium virial expansions) of cumulants of operator
groups of the von Neumann equations.

\textcolor{blue!50!black}{\subsection{Approaches to deriving quantum kinetic equations}}

Typically, when describing the evolution of many-particle quantum systems by kinetic equations,
it is assumed that the initial state is separable
\begin{eqnarray}\label{si}
    &&\hskip-7mm F^{(c)}(0)\equiv\big(I,F_1^{0}(1),\ldots,\otimes_{i=1}^n F_1^{0}(i),\ldots\big),
\end{eqnarray}
which means the absence of correlations between particles, or, in other words, a state satisfying
the chaos condition \cite{CGP97},\cite{G13}. We highlight that the assumption about the initial
state is a crucial hypothesis in current rigorous kinetic theory.

It should be observed that, for an initial state specified by the one-particle density operator
\eqref{si}, the Cauchy problem for the BBGKY hierarchy is not well-posed. Consequently, this
Cauchy problem can be reformulated as a new Cauchy problem for the one-particle density operator
and a sequence of explicitly defined functionals of the solution to the nonlinear evolution
equation for the one-particle density operator, known as the kinetic equation. The method for
constructing such state functionals relies on the application of kinetic cluster expansions \cite{GT12}.

The approach to describing the evolution of the state for separable initial states of quantum
many-body systems is based on sequences of reduced state functionals.

Indeed, we apply to the generating operators of the non-perturbative solution
$F(t)\equiv(I,F_1(t),F_2(t),$ $\ldots,F_s(t),\ldots)$ of the BBGKY hierarchy \eqref{dh}
\begin{eqnarray}\label{dhsi}
    &&\hskip-8mm F_{s}(t,1,\ldots,s)=\sum\limits_{n=0}^{\infty}\frac{1}{n!}\,\mathrm{Tr}_{s+1,\ldots,s+n}\,
       \mathfrak{A}_{1+n}(t,\{1,\ldots,s\},s+1,\ldots,s+n)\otimes_{i=1}^{s+n}F_1^{0}(i),\\
    &&\hskip-8mm s\geq1,\nonumber
\end{eqnarray}
the following cluster expansions, referred to as kinetic cluster expansions \cite{GT12}:
\begin{eqnarray}\label{kce}
   &&\hskip-7mm\mathfrak{A}_{1+n}(t,\{1,\ldots,s\},s+1,\ldots,s+n)=\sum_{n_1=0}^{n}\frac{n!}{(n-n_1)!}\,
     \mathfrak{V}_{1+n-n_1}(t,\{1,\ldots,s\},\\
   &&s+1,\ldots,s+n-n_1)
     \sum\limits_{\mbox{\scriptsize$\begin{array}{c}\mathrm{D}:(s+n-n_1+1,\ldots,s+n)=\bigcup_k X_k,\\|\mathrm{D}|
     \leq s+n-n_1\end{array}$}}\frac{1}{|\mathrm{D}|!}\times\nonumber\\
   &&\sum_{i_1\neq\ldots\neq i_{|\mathrm{D}|}=1}^{s+n-n_1}\,
     \prod_{X_{k}\subset \mathrm{D}}\,\frac{1}{|X_k|!}\,\mathfrak{A}_{1+|X_{k}|}(t,i_k,X_{k})
     \prod\limits_{\mbox{\scriptsize$\begin{array}{c}{m=1},
     \\m\neq i_1,\ldots,i_{|\mathrm{D}|}\end{array}$}}^{s+n-n_1}\mathfrak{A}_1(t,m),\quad n\geq0,\nonumber
\end{eqnarray}
where the symbol $\sum_{\mathrm{D}:(s+n-n_1+1,\ldots,s+n)=\bigcup_l X_l,\,|\mathrm{D}|\leq s+n-n_1}$ means
the sum over all possible dissections $\mathrm{D}$ of the linearly ordered set $(s+n-n_1+1,\ldots,s+n)$
on no more than $s+n-n_1$ linearly ordered subsets.
The first few of the reformulated cluster expansions \eqref{kce} take the form
\begin{eqnarray*}
   &&\hskip-7mm\widehat{\mathfrak{A}}_{1}(t,\{1,\ldots,s\})=\mathfrak{V}_{1}(t,\{1,\ldots,s\}),\\
   &&\hskip-7mm\widehat{\mathfrak{A}}_{2}(t,\{1,\ldots,s\},s+1)=\mathfrak{V}_{2}(t,\{1,\ldots,s\},s+1)+
       \mathfrak{V}_{1}(t,\{1,\ldots,s\})\sum_{i_1=1}^s \widehat{\mathfrak{A}}_{2}(t,i_1,s+1),\\
   && \vdots
\end{eqnarray*}

As a result, we establish that a sequence of reduced state functionals
$F(t\mid F_{1}(t))\equiv(I,F_1(t),F_2(t\mid F_{1}(t)),\ldots,F_s(t\mid F_{1}(t)),\ldots)$
is represented by following expansions into series:
\begin{eqnarray}\label{dhsi1}
    &&\hskip-9mm F_{1}(t,1)=\sum\limits_{n=0}^{\infty}\frac{1}{n!}\,\mathrm{Tr}_{2,\ldots,1+n}\,
       \mathfrak{A}_{1+n}(t,1,\ldots,1+n)\otimes_{i=1}^{1+n}F_1^{0}(i),\\
   \label{rf}
    &&\hskip-9mm F_{s}(t,1,\ldots,s\mid F_{1}(t))\doteq\\
    &&\sum_{n=0}^{\infty}\frac{1}{n!}\,\mathrm{Tr}_{s+1,\ldots,s+n}\,
        \mathfrak{V}_{1+n}(t,\{1,\ldots,s\},s+1,\ldots,s+n)\otimes_{i=1}^{s+n}F_{1}(t,i),\quad s\geq2.\nonumber
\end{eqnarray}
The generating operators of reduced state functionals \eqref{rf} are solutions of kinetic
cluster expansions:
\begin{eqnarray}\label{grf}
   &&\hskip-9mm\mathfrak{V}_{1+n}(t,\{1,\ldots,s\},s+1,\ldots,s+n)\doteq\\
   &&\hskip-5mm \sum_{k=0}^{n}\,(-1)^k\,\sum_{n_1=1}^{n}\ldots\sum_{n_k=1}^{n-n_1-\ldots-n_{k-1}}
      \frac{n!}{(n-n_1-\ldots-n_k)!}\times\nonumber\\
   &&\hskip-5mm\widehat{\mathfrak{A}}_{1+n-n_1-\ldots-n_k}(t,\{1,\ldots,s\},s+1,\ldots,s+n-n_1-\ldots-n_k)\times\nonumber\\
   &&\hskip-5mm\prod_{j=1}^k\sum\limits_{\mbox{\scriptsize$\begin{array}{c}\mathrm{D}_{j}:X_j=\bigcup_{l_j} Y_{l_j},\\
      |\mathrm{D}_{j}|\leq s+n-n_1-\dots-n_j\end{array}$}}\frac{1}{|\mathrm{D}_{j}|!}\sum_{i_1\neq\ldots\neq i_{|\mathrm{D}_{j}|}=1}^{s+n-n_1-\ldots-n_j}\prod_{Y_{l_j}\subset\mathrm{D}_{j}}\,\frac{1}{|Y_{l_j}|!}\,
      \widehat{\mathfrak{A}}_{1+|Y_{l_j}|}(t,i_{l_j},Y_{l_j}),\quad n\geq0,\nonumber
\end{eqnarray}
where, as above, we denoted $X_j\equiv(s+n-n_1-\ldots-n_j+1,\ldots,s+n-n_1-\ldots-n_{j-1})$, the symbol
$\sum_{\mathrm{D}:Y=\bigcup_l X_l,\,|\mathrm{D}|\leq s+n-n_1}$ means the sum over all possible dissections
$\mathrm{D}$ of the linearly ordered set $Y\equiv(s+n-n_1+1,\ldots,s+n)$ on no more than $s+n-n_1$ linearly
ordered subsets and the generating operator $\widehat{\mathfrak{A}}_{1+n}(t)$ of these expansions is the
$(1+n)$th-order cumulant of the groups of scattering operators \eqref{sc}.
Here are a few simple examples of the generating operators used in these expansions:
\begin{eqnarray*}
   &&\hskip-7mm \mathfrak{V}_{1}(t,\{1,\ldots,s\})=\widehat{\mathfrak{A}}_{1}(t,\{1,\ldots,s\}),\\
   &&\hskip-7mm \mathfrak{V}_{1+1}(t,\{1,\ldots,s\},s+1)=\widehat{\mathfrak{A}}_{2}(t,\{1,\ldots,s\},s+1)-
               \widehat{\mathfrak{A}}_{1}(t,\{1,\ldots,s\})\sum_{i=1}^s\widehat{\mathfrak{A}}_{2}(t,i,s+1),\\
   && \vdots
\end{eqnarray*}

Thus, the equality holds $F(t)=F(t\mid F_{1}(t))$, where the sequence $F(t)$ is a non-perturbative
solution \eqref{dhsi} of the BBGKY hierarchy \eqref{dh} for the initial data \eqref{si}, and the
sequence $F(t\mid F_{1}(t))$ is a sequence of reduced state functionals \eqref{rf}.
This relationship essentially elucidates the links among the equations that govern the state evolution
of many-particle quantum systems.


In the paper \cite{BG47}, Bogolyubov proposed a different approach to deriving Markovian kinetic equations.
This method relies on the construction of reduced state functionals $F_{s}\big(t,1,\ldots,s\mid F_{1}(t)\big)$
using perturbation theory methods. To compare, we briefly outline this approach.

First, consider the first terms of the series expansion \eqref{rf}. For a bounded interaction potential,
an analogue of the Duhamel equation holds for the second-order cumulant of the scattering operators \eqref{sc}
\begin{eqnarray*}\label{f}
    &&\hskip-7mm \widehat{\mathfrak{A}}_{2}(t,\{1,\ldots,s\},s+1)=
       \int_0^t d\tau\,\mathcal{G}^{\ast}_{s}(\tau,1,\ldots,s)
       \mathcal{G}^{\ast}_{1}(\tau,s+1)\sum\limits_{i_1=1}^{s}\mathcal{N}^{\ast}_{\mathrm{int}}(i_1,s+1)\times\\
    &&\widehat{\mathcal{G}}_{s+1}(t-\tau,1,\ldots,s+1)\prod_{i_2=1}^{s+1}\mathcal{G}^{\ast}_{1}(-\tau,i_2),
\end{eqnarray*}
and, consequently, the following representation holds for the second-order generating operator of series \eqref{rf}:
\begin{eqnarray*}\label{Df}
    &&\hskip-7mm \mathfrak{V}_{2}(t,\{1,\ldots,s\},s+1)\doteq\big(\widehat{\mathfrak{A}}_{2}(t,\{1,\ldots,s\},s+1)-
       \widehat{\mathfrak{A}}_{1}(t,\{1,\ldots,s\})\sum_{i_1=1}^s\widehat{\mathfrak{A}}_{2}(t,i_1,s+1)\big)=\\
    &&\int_0^t d\tau\,\mathcal{G}^{\ast}_{s}(\tau,1,\ldots,s)\mathcal{G}^{\ast}_{1}(\tau,s+1)\big(\sum\limits_{i_1=1}^{s}
       \mathcal{N}^{\ast}_{\mathrm{int}}(i_1,s+1))\widehat{\mathcal{G}}_{s+1}(\tau-t,1,\ldots,s+1)-\nonumber\\
    &&\widehat{\mathcal{G}}_{s}(\tau-t,1,\ldots,s)\sum\limits_{i_1=1}^{s}\mathcal{N}^{\ast}_{\mathrm{int}}(i_1,s+1)
       \widehat{\mathcal{G}}_{2}(\tau-t,i_1,s+1)\prod_{i_2=1}^{s+1}\mathcal{G}^{\ast}_{1}(-\tau,i_2).
\end{eqnarray*}
The characteristic scale for kinetic phenomena described, for example, by the quantum Boltzmann equation is determined
by the mean free path. On the kinetic scale, the variation of the operator group variables \eqref{dgroup} for a finite
number of particles depends on the microscopic time variable $\varepsilon^{-1}t$, where $\varepsilon\geq0$ is a scaling
parameter, and the dimensionless reduced state functionals are represented in the form:
$F_{s}(\varepsilon^{-1}t,1,\ldots,s\mid F_{1}(t))$.
Then, according to the Duhamel equation for the second-order cumulant of scattering operators, in the formal Markovian
limit $\varepsilon \to 0$, the first two terms of the expansions of the dimensionless reduced state functionals coincide
with the corresponding terms obtained via perturbation theory using the correlation weakening condition in \cite{BG47}
\begin{eqnarray*}
    &&\hskip-7mm \lim\limits_{\epsilon\rightarrow 0}F_{s}(\varepsilon^{-1}t,1,\ldots,s\mid F_{1}(t))=
       \widehat{\mathcal{G}}_{s}(\infty,1,\ldots,s)\prod_{i=1}^{s}F_{1}(t,i)+\\
    &&\int_0^{\infty} d\tau\,\mathcal{G}^{\ast}_{s}(\tau,1,\ldots,s)\mathrm{Tr}_{s+1}\big(\sum\limits_{i_1=1}^{s}
       \mathcal{N}^{\ast}_{\mathrm{int}}(i_1,s+1)\widehat{\mathcal{G}}_{s+1}(\infty,1,\ldots,s,s+1)-\nonumber\\
    &&\widehat{\mathcal{G}}_{s}(\infty,1,\ldots,s)\sum\limits_{i_1=1}^{s}\mathcal{N}^{\ast}_{\mathrm{int}}(i_1,s+1)
       \widehat{\mathcal{G}}_{2}(\infty,i_1,s+1)\big)\prod_{i_2=1}^{s+1}\mathcal{G}^{\ast}_{1}(-\tau,i_2)F_{1}(t,i_2)+etc.
\end{eqnarray*}

Thus, at the specified kinetic scale, the collision integral of the generalized kinetic equation \eqref{gke}
takes the form of the Bogolyubov collision integral \cite{BG47}. It should be noted that, in the homogeneous
case, the collision integral in the first approximation of the resulting expression has a more general form
than the collision integral of the quantum Boltzmann equation.


\textcolor{blue!50!black}{\subsection{Evolution of correlations governed by the kinetic equation}}

Let us formulate the evolution equation for the one-particle reduced density operator representing
by series \eqref{dhsi1}.
By differentiating the series expansion \eqref{dhsi1} with respect to the time variable in the sense
of pointwise convergence in the space $\mathfrak{L}^{1}(\mathcal{H})$ and applying the kinetic cluster
expansions \eqref{kce} for the cumulants of operator groups \eqref{drcum}, we derive an identity for
the one-particle density operator. The resulting identity for the one-particle density operator is
then treated as an evolution equation governing the one-particle state of quantum systems.

If $F_{1}^{0}\in\mathfrak{L}^{1}(\mathcal{H})$, then for an arbitrary $t\in\mathbb{R}$, the one-particle
reduced density operator \eqref{dhsi1} is governed by the Cauchy problem for the following generalized
quantum kinetic equation \cite{GT12}:
\begin{eqnarray}\label{gke}
    &&\hskip-7mm \frac{d}{dt}F_{1}(t,1)=\mathcal{N}_{1}^{\ast}(1)F_{1}(t,1)+
       \mathrm{Tr}_{2}\,\mathcal{N}_{\mathrm{int}}^{\ast}(1,2)F_{2}(t,1,2\mid F_{1}(t)),\\
    \label{gkei}
    &&\hskip-7mm F_1(t,1)|_{t=0}= F_1^{0}(1),
\end{eqnarray}
where the collision integral in this kinetic equation is determined by the two-particle reduced functional
of the state \eqref{rf}, and the operators $\mathcal{N}^{\ast}_{1}$ and $\mathcal{N}^{\ast}_{\mathrm{int}}$
are defined by the formula \eqref{vNeg}.

In the space of trace class operators $\mathfrak{L}^{1}(\mathcal{H})$ the following statement holds \cite{GT12}.

\medskip
\begin{theorem}
For bounded interaction potentials, the global-in-time non-perturbative solution to the Cauchy problem
\eqref{gke},\eqref{gkei} is determined by a one-particle reduced density operator represented by the
series expansion \eqref{dhsi1}.
If $\|F_1^0\|_{\mathfrak{L}^{1}(\mathcal{H})}<e^{-2}$, then, provided
$F_1^0\in\mathfrak{L}^{1}_{0}(\mathcal{H})$, it is a strong (classical) solution, whereas for arbitrary
initial data $F_1^{0}\in\mathfrak{L}^{1}(\mathcal{H})$, it is a weak (generalized) solution.
\end{theorem}

The outline of the proof for this assertion is analogous to the proof of the existence theorem for
the BBGKY hierarchy \eqref{dh} in the space of sequences of trace-class operators.

Thus, for a separable initial state \eqref{si}, the evolution of the state of a system of many quantum
particles is described in terms of the state of a typical particle governed by a non-Markovian nonlinear
kinetic equation \eqref{gke}, and a sequence of functionals \eqref{rf} of this state \eqref{dhsi1}
without any scaling approximations. The state functionals \eqref{rf} describe the process of creating
correlations in a quantum system of particles by employing the kinetic equation \eqref{gke}.

\textcolor{blue!50!black}{\subsection{The process of creating correlations from chaos}}

To describe the process of correlation propagation for the separable initial state \eqref{si},
we use the reduced correlation operators \eqref{Gg}, defined in terms of the correlation operators
\eqref{cch}. As a result, the reduced correlation functionals are represented by the following series
expansions:
\begin{eqnarray}\label{fch}
    &&\hskip-7mm G_{s}(t,1,\ldots,s\mid G_{1}(t))\doteq\sum_{n=0}^{\infty}\frac{1}{n!}\,
      \mathrm{Tr}_{s+1,\ldots,s+n}\,g_{s+n}(t,1,\ldots,s+n\mid g_{1}(t))=\\
    &&\sum_{n=0}^{\infty}\frac{1}{n!}\,
      \mathrm{Tr}_{s+1,\ldots,s+n}\,\mathfrak{V}_{s+n}(t,1,\ldots,s+n)\otimes_{j=1}^{s+n}G_{1}(t,j),
      \quad s\geq2,\nonumber
\end{eqnarray}
where the generating operators $\mathfrak{V}_{s+n}(t),\,n\geq0,$ are $(s+n)$th-order operators represented
by the expansions \eqref{grf} for $s+n$ individual particles and
\begin{eqnarray*}
   &&\hskip-7mm G_{1}(t,1)=\sum\limits_{n=0}^{\infty}\frac{1}{n!}\,\mathrm{Tr}_{2,\ldots,1+n}\,
      \mathfrak{A}_{1+n}(t,1,\ldots,1+n)\,\otimes_{j=1}^{1+n}G_{1}^{0}(j).
\end{eqnarray*}
The sequence of reduced correlation functionals \eqref{fch} describes the process of creating correlations
from chaos in a quantum system with infinitely many particles.

It should be emphasized that the reduced correlation functionals \eqref{fch} are cumulants of the reduced
state functionals \eqref{rf}. The generating operators for the series expansions \eqref{fch} and \eqref{rf}
differ in their order.
The relationship between the reduced correlation functionals \eqref{fch} and reduced state functionals
\eqref{rf} elucidates the links among the equations that govern the state evolution of many-particle
quantum systems.

The reduced correlation functionals \eqref{fch} on the microscopic scale directly determine the macroscopic
characteristics of fluctuations of observables. For example, the functional representing the dispersion of
an additive-type observable, i.e., the sequence $O^{(1)}=(0,o_{1}(1),\ldots,\sum_{i_{1}=1}^{n}o_1(i_{1}),\ldots)$,
is given by the formula
\begin{eqnarray*}
    &&\hskip-7mm \langle(O^{(1)}-\langle O^{(1)}\rangle)^2\rangle(t)=\\
    &&\mathrm{Tr}_{1}\,(o_1^2(1)- \langle O^{(1)}\rangle^2(t))G_{1}(t,1)+
       \mathrm{Tr}_{1,2}\,o_{1}(1)o_{1}(2)G_{2}(t,1,2\mid G_{1}(t)),
\end{eqnarray*}
where $\langle O^{(1)}\rangle(t)=\mathrm{Tr}_{1}\,o_{1}(1)G_{1}(t,1)$ is the mean-value functional of
an additive-type observable.

\textcolor{blue!50!black}{\subsection{The mean-field scaling approximation for a separable initial state}}

The widely accepted approach to describing evolution through kinetic equations is as follows
\cite{CGP97},\cite{P2021}. If the initial state of a system composed of many particles is
specified by the state of a typical particle of the system, meaning there are no correlations
between the particles at the initial instant, then under certain scaling approximations, the
evolution of the system can be effectively described as the evolution of a typical particle
state. This approach is formulated in terms of a one-particle reduced density operator, which
is governed by a Markovian nonlinear kinetic equation.

The specific quantum kinetic equations, such as the quantum Boltzmann equation and others,
can be derived from the generalized quantum kinetic equation \eqref{gke} under appropriate
scaling approximations.
Below, we will argue the asymptotic behavior in the mean-field scaling limit of the solution
\eqref{dhsi} of the generalized quantum kinetic equation \eqref{gke} and the sequence of
reduced state functionals \eqref{rf}.

To determine the scaling parameter, we rewrite the kinetic equation \eqref{gke} in dimensionless
form. Then the collision integral of this kinetic equation takes the form:
$\epsilon\,\mathrm{Tr}_{2}\,\mathcal{N}_{\mathrm{int}}^{\ast}(1,2)F_{2}(t,1,2\mid F_{1}(t))$,
where $\epsilon > 0$, is a scaling parameter, which is a consequence of the dimensionless form
of the Hamiltonian
$H_{n}={\sum\limits}_{j=1}^{n}K(j)+\epsilon\,{\sum\limits}_{j_{1}<j_{2}=1}^{n}\Phi(j_{1},j_{2})$.

Since for an arbitrary finite time interval for an asymptotically perturbed first-order cumulant of
strongly continuous operator groups \eqref{dgroup} such equality holds
\begin{eqnarray}\label{apgc1}
    &&\hskip-8mm\lim\limits_{\epsilon\rightarrow 0}\big\|\mathcal{G}^{\ast}_{s}(t,1,\ldots,s)f_{s}-
        \prod\limits_{j=1}^{s}\mathcal{G}^{\ast}_{1}(t,j)f_{s}\big\|_{\mathfrak{L}^{1}(\mathcal{H}^{\otimes s})}=0.
\end{eqnarray}
then for the $(1+n)$th-order cumulant of asymptotically perturbed operator groups \eqref{dgroup}
the equality holds
\begin{eqnarray}\label{apgc}
    &&\hskip-8mm\lim\limits_{\epsilon\rightarrow 0}\big\|\frac{1}{\epsilon^n}\,
       \frac{1}{n!}\mathfrak{A}_{1+n}(t,\{1,\ldots,s\},s+1,\ldots,s+n)f_{s+n}-\\
    &&\int\limits_0^tdt_{1}\ldots\int\limits_0^{t_{n-1}}dt_{n}
        \prod\limits_{j=1}^{s}\mathcal{G}_{1}^{\ast}(t-t_{1},j)
        \sum\limits_{i_{1}=1}^{s}\mathcal{N}^{\ast}_{\mathrm{int}}(i_{1},s+1)
        \prod\limits_{j_1=1}^{s+1}\mathcal{G}_{1}^{\ast}(t_{1}-t_{2},j_1)\ldots\nonumber\\
    &&\prod\limits_{j_{n-1}=1}^{s+n-1}\mathcal{G}^{\ast}_{1}(t_{n-1}-t_{n},j_{n-1})
        \sum\limits_{i_{n}=1}^{s+n-1}\mathcal{N}^{\ast}_{\mathrm{int}}(i_{n},s+n)
        \prod\limits_{j_n=1}^{s+n}\mathcal{G}^{\ast}_{1}(t_{n},j_n)
        f_{s+n}\big\|_{\mathfrak{L}^{1}(\mathcal{H}^{\otimes s+n})}=0,\nonumber\\
    &&\hskip-8mm n\geq1.\nonumber
\end{eqnarray}
The last equalities are a consequence of the validity for the bounded interaction potential of analogues
of the Duhamel equation for cumulants \eqref{dcum} of asymptotically perturbed operator groups \eqref{dgroup}.

According to equalities \eqref{apgc1} and \eqref{apgc} for a one-particle reduced density operator \eqref{dhsi1}
the following limit theorem holds \cite{GT12}.

\medskip
\begin{theorem}
Let the mean-field limit exist for the initial one-particle reduced density operator in the following sense:
\begin{eqnarray*}\label{asic1}
   &&\hskip-8mm\lim\limits_{\epsilon\rightarrow 0}\big\|\epsilon\,F_{1}^{0}-
      f_{1}^{0}\big\|_{\mathfrak{L}^{1}(\mathcal{H})}=0,
\end{eqnarray*}
then for an arbitrary finite time interval for series expansion \eqref{dhsi1} the equality holds
\begin{eqnarray*}
   &&\hskip-8mm \lim\limits_{\epsilon\rightarrow 0}\big\|\epsilon\, F_{1}(t)-
     f_{1}(t)\big\|_{\mathfrak{L}^{1}(\mathcal{H})}=0,
\end{eqnarray*}
where the limiting one-particle density operator $f_1(t)$ is represented by a series converging in the norm
of the space $\mathfrak{L}^{1}(\mathcal{H})$
\begin{eqnarray}\label{viterc}
     &&\hskip-8mm f_{1}(t,1)=\sum\limits_{n=0}^{\infty}\,\int\limits_0^tdt_{1}\ldots
        \int\limits_0^{t_{n-1}}dt_{n}\,\mathrm{Tr}_{2,\ldots,n+1}\mathcal{G}^{\ast}_{1}(t-t_{1},1)
        \mathcal{N}^{\ast}_{\mathrm{int}}(1,2)\prod\limits_{j_1=1}^{2}\mathcal{G}^{\ast}_{1}(t_{1}-t_{2},j_1)\ldots\\
     &&\prod\limits_{i_{n}=1}^{n}\mathcal{G}^{\ast}_{1}(t_{n-1}-t_{n},i_{n})
         \sum\limits_{k_{n}=1}^{n}\mathcal{N}^{\ast}_{\mathrm{int}}(k_{n},n+1)
         \prod\limits_{j_n=1}^{n+1}\mathcal{G}^{\ast}_{1}(t_{n},j_n)\otimes_{i=1}^{1+n}f_1^0(i).\nonumber
\end{eqnarray}
\end{theorem}

For bounded interaction potentials, the series \eqref{viterc} converges in the norm of the space
$\mathfrak{L}^{1}(\mathcal{H})$ provided that
$t<t_{0}\equiv(2\,\|\Phi\|_{\mathfrak{L}(\mathcal{H}^{\otimes 2})}\|f_1^0\|_{\mathfrak{L}^{1}(\mathcal{H})})^{-1}$.

For the initial state $f_{1}^0\in\mathfrak{L}^{1}(\mathcal{H})$ the limiting operator \eqref{viterc} satisfies
the Cauchy problem for the quantum Vlasov kinetic equation \cite{G13}
\begin{eqnarray}\label{Vlasov1}
     &&\hskip-7mm \frac{\partial}{\partial t}f_{1}(t,1)=\mathcal{N}^{\ast}_{1}(1)f_{1}(t,1)+
       \mathrm{Tr}_{2}\,\mathcal{N}^{\ast}_{\mathrm{int}}(1,2)f_{1}(t,1)\otimes f_{1}(t,2),\\
   \label{Vlasovi}
     &&\hskip-7mm f_{1}(t,1)|_{t=0}=f_{1}^0(1).
\end{eqnarray}

Note that for pure states—that is, for $f_{1}(t)=|\psi(t)\rangle\langle\psi(t)|$ or, in terms 
of the kernel of such an operator, $f_{1}(t,q,q')=\psi(t,q)\psi^{\ast}(t,q')$—the quantum Vlasov 
kinetic equation \eqref{Vlasov1} reduces to the Hartree equation
\begin{eqnarray*}
    &&\hskip-7mm i\frac{\partial}{\partial t} \psi(t,q)=-\frac{1}{2}\Delta_{q}\psi(t,q)+
       \int dq'\Phi(q-q')|\psi(q')|^{2}\psi(t,q).
\end{eqnarray*}
In particular, if the interaction potential kernel $\Phi(q)=\delta(q)$ is a Dirac measure, then
the Hartree kinetic equation reduces to the nonlinear Schr\"{o}dinger equation with cubic nonlinearity.
\begin{eqnarray*}
    &&i\hskip-7mm \frac{\partial}{\partial t}\psi(t,q)=
       -\frac{1}{2}\Delta_{q}\psi(t,q)+ |\psi(t,q)|^{2}\psi(t,q).
\end{eqnarray*}

By virtue of property \eqref{apgc} of the cumulants of asymptotically perturbed operator groups, the
following equalities hold for the generating operators \eqref{grf} of reduced state functionals \eqref{rf}:
\begin{eqnarray*}
    &&\hskip-7mm \lim\limits_{\epsilon\rightarrow 0}\big\|\big(\mathfrak{V}_{1}(t,\{1,\ldots,s\})-I\big)f_{s}
         \big\|_{\mathfrak{L}^{1}(\mathcal{H}^{\otimes s})}=0,
\end{eqnarray*}
and
\begin{eqnarray*}
    &&\hskip-7mm \lim\limits_{\epsilon\rightarrow 0}\big\|\frac{1}{\epsilon^{n}}\,
        \mathfrak{V}_{1+n}(t,\{1,\ldots,s\},s+1,\ldots,s+n)f_{s+n}
        \big\|_{\mathfrak{L}^{1}(\mathcal{H}^{\otimes s+n})}=0, \quad n\geq1,
\end{eqnarray*}
respectively.
Due to such equalities and convergence in the mean-field limit of the solution of the Cauchy problem of the
generalized quantum kinetic equation \eqref{gke},\eqref{gkei} to the solution of the Cauchy problem of the
Vlasov quantum kinetic equation \eqref{Vlasov1},\eqref{Vlasovi} for reduced state functionals \eqref{rf}
the following limit theorem holds.

\medskip
\begin{theorem}
For the reduced state functionals \eqref{rf}, the following equalities hold:
\begin{eqnarray*}
    &&\hskip-7mm \lim\limits_{\epsilon\rightarrow 0}\big\|\epsilon^{s} F_{s}\big(t,1,\ldots,s \mid F_{1}(t)\big)-
       \otimes_{j=1}^{s}f_{1}(t,j)\big\|_{\mathfrak{L}^{1}(\mathcal{H}^{\otimes s})}=0, \quad s\geq2,
\end{eqnarray*}
where the limit operator $f_{1}(t)$ is determined by series \eqref{viterc}, which represents the perturbative
solution of the quantum Vlasov kinetic equation \eqref{Vlasov1}.
\end{theorem}

The last assertion describes how initial chaos propagates in the mean-field limit. If the system is initially
uncorrelated, then no correlations arise during its evolution within this approximation.

It should be noted that the traditional approach to the problem of the propagation of initial chaos
is based on constructing the asymptotic behavior of the solution to the quantum BBGKY hierarchy \eqref{dh}
for reduced density operators within the framework of perturbation theory \cite{G13},\cite{BPS}.

\textcolor{blue!50!black}{\section{Kinetic evolution founded on observables}}

This section will clarify the meaning of kinetic evolution in terms of observables for many quantum particles.
A rigorous description of the kinetic evolution in terms of observables will be achieved by examining the
mean-field asymptotics of the non-perturbative solution \eqref{sroh} to the Cauchy problem for the hierarchy
of evolution equations for observables \eqref{roh}. We will also establish how the evolution described by
observables for a system of many quantum particles in the scaling approximation is related to the kinetic
evolution of the state of a typical particle in that system.

Note that, for the separable initial state \eqref{si}, the kinetic representation of the mean-value
functional for observables \eqref{rev} holds
\begin{eqnarray}\label{BF_1}
    &&\big(B(t),F^{(c)}\big)=\big(B(0),F(t\mid F_{1}(t))\big),
\end{eqnarray}
where the sequence $F(t\mid F_{1}(t))$ of reduced state functionals defined by formulas \eqref{dhsi1}
 and \eqref{rf}.
For additive-type reduced observables $B^{(1)}(0)=(0,b_{1}(1),0,\ldots)$ the mean-value functional for
observables \eqref{BF_1} takes the form
\begin{eqnarray*}
    &&\big(B^{(1)}(t),F^{(c)}\big)=\mathrm{Tr}_{1}\,b_{1}(1)F_{1}(t),
\end{eqnarray*}
where the one-particle reduced density operator $F_{1}(t)$ is given by the series \eqref{dhsi1}.
For $k$-ary reduced observables $B^{(k)}(0)=(0,\ldots,0,b_{k}(1,\ldots,k),0,\ldots),\,k\geq 2,$
the kinetic representation \eqref{BF_1} takes the form
\begin{eqnarray*}
    &&\big(B^{(k)}(t),F^{(c)}\big)=
       \frac{1}{k!}\, \mathrm{Tr}_{1,\ldots,k}\,b_{k}(1,\ldots,k)F_k(t,1,\ldots,k\mid F_{1}(t)),
\end{eqnarray*}
where the reduced state functional $F_k(t,1,\ldots,k\mid F_{1}(t))$ is given by the series \eqref{rf}.

\textcolor{blue!50!black}{\subsection{The hierarchy of quantum kinetic equations for observables}}

To determine the scaling parameter, we rewrite the hierarchy of evolution equations \eqref{roh} in dimensionless
form, as before. Then the generator of this hierarchy takes the form:
\begin{eqnarray*}
    &&\hskip-7mm (\mathcal{N}_{s}b)_{s}(1,\ldots,s)=
     \sum_{j=1}^{s}\mathcal{N}(j)+\epsilon\,\sum_{j_1<j_{2}=1}^{s}\mathcal{N}_{\mathrm{int}}(j_{1},j_{2}),
\end{eqnarray*}
\begin{eqnarray*}
    &&\hskip-7mm (\big[\mathcal{N},\mathfrak{a}^+\big]b)_{s}(1,\ldots,s)
       =\epsilon\,\sum_{j_1\neq j_{2}=1}^s\mathcal{N}_{\mathrm{int}}(j_1,j_{2})b_{s-1}((1,\ldots,s)\setminus j_1)
       \otimes \mathbbm{1}_{(j_1)},\quad s\geq 2.
\end{eqnarray*}
where $\epsilon > 0$ is a scaling parameter, which is a consequence of the dimensionless form
of the Hamiltonian.

Suppose that at the initial moment there exists the mean-field limit of reduced observables \eqref{rohi}
in the sense of $\ast$-weak convergence in the space of bounded operators $\mathfrak{L}(\mathcal{H}^{\otimes s})$
\begin{eqnarray}\label{asumdino}
    &&\mathrm{w^{\ast}-}\lim\limits_{\epsilon\rightarrow 0}\big(\epsilon^{-s}B_{s}^{\epsilon,0}-b_{s}^0\big)=0,
\end{eqnarray}
where $\epsilon>0$ is a scaling parameter.
Then the following limit theorem holds for the non-perturbative solution \eqref{sroh} of the hierarchy of
evolution equations \eqref{roh}.

\medskip
\begin{theorem}
If condition \eqref{asumdino} is satisfied, then for a finite time interval, the mean-field limit exists
in the same sense for the sequence of reduced observables \eqref{sroh} as well \cite{G11},\cite{G15}
\begin{eqnarray*}\label{asymto}
    &&\mathrm{w^{\ast}-}\lim\limits_{\epsilon\rightarrow 0}\big(\epsilon^{-s}B_{s}(t)-b_{s}(t)\big)=0,
       \quad  s\geq1,
\end{eqnarray*}
where the limiting reduced observables $b_{s}(t),\,s\geq1$, are defined by the following expansions:
\begin{eqnarray}\label{Iterd}
   &&\hskip-8mm  b_{s}(t,1,\ldots,s)=\sum\limits_{n=0}^{s-1}\,
      \int\limits_0^tdt_{1}\ldots\int\limits_0^{t_{n-1}}dt_{n}
      \,\mathcal{G}_s^{0}(t-t_{1})\sum\limits_{i_{1}\neq j_{1}=1}^{s}
      \mathcal{N}_{\mathrm{int}}(i_{1},j_{1})\mathcal{G}_{s-1}^{0}(t_{1}-t_{2})\ldots\\
   &&\mathcal{G}_{s-n+1}^{0}(t_{n-1}-t_{n})
      \sum\limits^{s}_{\mbox{\scriptsize $\begin{array}{c}i_{n}\neq j_{n}=1,\\
      i_{n},j_{n}\neq(j_{1},\ldots,j_{n-1})\end{array}$}}\mathcal{N}_{\mathrm{int}}(i_{n},j_{n})
      \mathcal{G}_{s-n}^{0}(t_{n})\times\nonumber\\
   && b_{s-n}^0((1,\ldots,s)\setminus({j_{1}},\ldots,{j_{n}}))\otimes
      \mathbbm{1}_{(j_1)}\ldots\otimes\mathbbm{1}_{(j_n)},\quad  s\geq1,\nonumber
\end{eqnarray}
and for the operator groups of non-interacting particles, the notation was used
\begin{eqnarray*}
   && \mathcal{G}_{s-n+1}^{0}(t_{n-1}-t_{n})\equiv
      \prod\limits_{j\in (1,\ldots,s)\setminus (j_1,\ldots,j_{n-1})}\mathcal{G}_{1}(t_{n-1}-t_{n},j).
\end{eqnarray*}
\end{theorem}

For a certain type of reduced observable, the structure of expansions \eqref{Iterd} takes a specific form.
For example, in the case of reduced observables of $k$-ary type, we have:
\begin{eqnarray}\label{kIterd}
   &&\hskip-8mm  b_{s}^{(k)}(t,1,\ldots,s)=\int\limits_0^tdt_{1}\ldots\int\limits_0^{t_{s-k-1}}dt_{s-k}
      \,\mathcal{G}_s^{0}(t-t_{1})\sum\limits_{i_{1}\neq j_{1}=1}^{s}\mathcal{N}_{\mathrm{int}}(i_{1},j_{1})
      \mathcal{G}_{s-1}^{0}(t_{1}-t_{2})\ldots\\
   && \mathcal{G}_{s-n+1}^{0}(t_{s-k-1}-t_{s-k})\sum\limits^{s}_{\mbox{\scriptsize $\begin{array}{c}i_{s-k}\neq j_{s-k}=1,\\
      i_{s-k},j_{s-k}\neq (j_{1},\ldots,j_{s-k-1})\end{array}$}}\mathcal{N}_{\mathrm{int}}(i_{s-k},j_{s-k})
      \mathcal{G}_{s-n}^{0}(t_{s-k})\times\nonumber\\
   &&  b_{k}^0((1,\ldots,s)\setminus({j_{1}},\ldots,{j_{s-k}}))\otimes\mathbbm{1}_{(j_1)}\ldots\otimes\mathbbm{1}_{(j_{s-k})},
   \quad  1\leq s\leq k.\nonumber
\end{eqnarray}

In the space $b^0\in\mathfrak{L}_\gamma(\mathcal{F}_{\mathcal{H}})$ the sequence
$b(t)=(b_0,b_1(t),\ldots,b_{s}(t),\ldots)$ of limiting reduced observables \eqref{Iterd} satisfies
the Cauchy problem for the so-called dual Vlasov hierarchy:
\begin{eqnarray}\label{vdh}
   &&\hskip-8mm \frac{\partial}{\partial t}b_{s}(t,1,\ldots,s)=
      \sum\limits_{j=1}^{s}\mathcal{N}(j)\,b_{s}(t,1,\ldots,s)+\\
   && \sum_{j_1\neq j_{2}=1}^s\mathcal{N}_{\mathrm{int}}(j_1,j_{2})\,
     b_{s-1}(t,(1,\ldots,s)\setminus(j_1))\otimes \mathbbm{1}_{(j_1)},\nonumber\\
    \label{vdhi}
   &&\hskip-8mm b_{s}(t,1,\ldots,s)|_{t=0}=b_{s}^0(1,\ldots,s),\quad s\geq1.
\end{eqnarray}
It should be noted that the sequence of evolution equations \eqref{vdh} has the structure
of recurrence evolution equations.

Let us give a few examples of equations of the dual Vlasov hierarchy \eqref{vdh} in terms of
the kernels of operators to the limiting reduced observables:
\begin{eqnarray*}
    &&\hskip-8mm i\,\frac{\partial}{\partial t}b_{1}(t,q_1;q'_1)=
       -\frac{1}{2}(-\Delta_{q_1}+\Delta_{q'_1})b_{1}(t,q_1;q'_1),\\
    &&\hskip-8mm i\,\frac{\partial}{\partial t}b_{2}(t,q_1,q_2;q'_1,q'_2)=\\
    &&\big(-\frac{1}{2}\sum\limits_{i=1}^2(-\Delta_{q_i}+\Delta_{q'_i})+
       (\Phi(q'_1-q'_2)-\Phi(q_1-q_2))\big)b_{2}(t,q_1,q_2;q'_1,q'_2))+\\
    &&\big(\Phi(q'_1-q'_2)-\Phi(q_1-q_2)\big)\big(b_{1}(t,q_1;q'_1)+b_{1}(t,q_2;q'_2)\big),\\
    &&\vdots
\end{eqnarray*}

Thus, in the mean-field limit, the kinetic behavior of quantum systems of many particles
is described in terms of the sequence of limiting reduced observables \eqref{Iterd},
in particular, by the sequence \eqref{kIterd}, whose evolution is governed by the Cauchy
problem of the dual Vlasov hierarchy \eqref{vdh},\eqref{vdhi}.

\textcolor{blue!50!black}{\subsection{The propagation of the initial chaos}}

Let us consider the relationship between the kinetic behavior within the mean-field approximation
for a many-particle quantum system described by the dual Vlasov hierarchy \eqref{vdh} for limiting
reduced observables, and the Vlasov kinetic equation \eqref{Vlasov1} for the state of a typical
particle of the system.

Let the initial state be separable, i.e., it is specified by such a sequence of limiting reduced
density operators
\begin{eqnarray}\label{h}
      &&\hskip-7mm f^{(c)}\equiv\big(I,f_1^{0}(1),\ldots,\otimes_{i=1}^s f_1^{0}(i),\ldots\big).
\end{eqnarray}
As previously noted, this assumption regarding the initial state is characteristic of the kinetic
description of a gas, since in this case the state is completely determined by the one-particle
density operator.

For the mean-value functional for limiting reduced additive-type observables and a separable
initial state, the following equality holds on a finite time interval:
\begin{eqnarray*}
  &&\hskip-8mm\big(b^{(1)}(t),f^{(c)}\big)=\sum\limits_{s=0}^{\infty}\frac{1}{s!}\,
         \mathrm{Tr}_{1,\ldots,s}\,b_{s}^{(1)}(t,1,\ldots,s)\otimes_{i=1}^{s}f_{1}^0(i)=\\
  &&\mathrm{Tr}_{1}\,b_{1}^{0}(1)f_{1}(t,1),\nonumber
\end{eqnarray*}
where the limiting one-particle density operator is represented as a perturbation theory series \eqref{viterc}
\begin{eqnarray}\label{viter}
     &&\hskip-8mm f_{1}(t,1)=\sum\limits_{n=0}^{\infty}\,\int\limits_0^tdt_{1}\ldots
        \int\limits_0^{t_{n-1}}dt_{n}\,\mathrm{Tr}_{2,\ldots,n+1}\mathcal{G}^{\ast}_{1}(t-t_{1},1)
        \mathcal{N}^{\ast}_{\mathrm{int}}(1,2)\prod\limits_{j_1=1}^{2}\mathcal{G}^{\ast}_{1}(t_{1}-t_{2},j_1)\ldots\\
     &&\prod\limits_{i_{n}=1}^{n}\mathcal{G}^{\ast}_{1}(t_{n-1}-t_{n},i_{n})
         \sum\limits_{k_{n}=1}^{n}\mathcal{N}^{\ast}_{\mathrm{int}}(k_{n},n+1)
         \prod\limits_{j_n=1}^{n+1}\mathcal{G}^{\ast}_{1}(t_{n},j_n)\otimes_{i=1}^{1+n}f_1^0(i).\nonumber
\end{eqnarray}
The one-particle density operator represented by the series \eqref{viter} satisfies the quantum
Vlasov kinetic equation \eqref{Vlasov1}.

Thus, the hierarchy of evolution equations \eqref{vdh} for limiting reduced observables of the
additive-type and for a separable initial state \eqref{h} describes the evolution of quantum
many-particle systems in a way equivalent to the Vlasov quantum kinetic equation \eqref{Vlasov1}.

Accordingly, for the mean-value functional of the limiting reduced observables of nonadditive-type
\eqref{kIterd} and a separable initial state \eqref{h}, the following equality holds on a finite
time interval:
\begin{eqnarray}\label{pchaos}
    &&\hskip-7mm \big(b^{(k)}(t),f^{(c)}\big)=\sum\limits_{s=0}^{\infty}\frac{1}{s!}\,
         \mathrm{Tr}_{1,\ldots,s}\,b_{s}^{(k)}(t,1,\ldots,s)\prod\limits_{i=1}^{s}f_1^0(i)=\\
    &&\frac{1}{k!}\mathrm{Tr}_{1,\ldots,k}\,b_{k}^{0}(1,\ldots,k)
       \otimes_{i=1}^{k}f_{1}(t,i),\quad k\geq2,\nonumber
\end{eqnarray}
where the one-particle reduced density operator is represented as a perturbation theory series
\eqref{viter}, i.e., satisfies the Cauchy problem for the Vlasov quantum kinetic equation
\eqref{Vlasov1},\eqref{Vlasovi}.
Equality \eqref{pchaos} describes the propagation of initial chaos \eqref{h} in the mean-field limit
via the solution \eqref{viter} to the quantum Vlasov kinetic equation \eqref{Vlasov1}. Specifically,
for any finite time interval, the state is represented by a sequence of limiting reduced operators:
\begin{eqnarray}\label{ihaos}
     &&\hskip-7mm f_k(t,1,\ldots,k)=\otimes_{i=1}^{k}f_{1}(t,i),\quad k\geq2.
\end{eqnarray}
That is, if there are no correlations in the particle system at the initial time, then within this
approximation, correlations between particle states do not arise during the evolution process, and
individual particle states evolve according to the kinetic equation \eqref{Vlasov1}.

\textcolor{blue!50!black}{\subsection{On the propagation of initial correlations}}

It should be noted that the approach to deriving kinetic equations discussed above allows for their
formulation for more general initial states, describing not only gases of quantum particles but
also many-particle systems in condensed states.

Let the initial state of a quantum system of many particles be specified by a one-particle density
operator and correlation operators \cite{G15}
\begin{eqnarray}\label{isc}
   &&\hskip-7mm f^{(cc)}\equiv(I,f_1^{0}(1),\otimes_{i=1}^{2}f_1^{0}(i)\otimes g_{2}^0(1,2),
        \ldots,\otimes_{i=1}^{n}f_1^{0}(i)\otimes g_{n}^0(1,\ldots,n),\ldots),
\end{eqnarray}
where the correlations of the initial states of the particles are specified by the operators
$g_{n}^0\in\mathfrak{L}^{1}_0(\mathcal{H}^{\otimes n}),$  $n\geq2$. It should be emphasized that
this assumption \eqref{isc} regarding the initial state is characteristic of the kinetic description
of many-particle systems in condensed states, which are inherently marked by the presence of correlations.

Then, using a method for deriving kinetic equations based on the hierarchy of kinetic equations
for reduced observables, we establish that, for an arbitrary finite time interval, the state is
described by a sequence $f(t)=(I,f_{1}(t),\ldots,f_{n}(t,1,\ldots,n),\ldots)$ of limiting reduced
density operators, with the one-particle density operator being represented by a corresponding
perturbation theory series expansion
\begin{eqnarray}\label{vitercc}
     &&\hskip-7mm f_{1}(t,1)=\\
     &&\sum\limits_{n=0}^{\infty}
        \int\limits_0^tdt_{1}\ldots\int\limits_0^{t_{n-1}}dt_{n}\,
        \mathrm{Tr}_{2,\ldots,n+1}\prod\limits_{i_1=1}^{1}\mathcal{G}^{\ast}_{1}(t-t_{1},i_1)
        \mathcal{N}^{\ast}_{\mathrm{int}}(1,2)\prod\limits_{j_1=1}^{2}
        \mathcal{G}^{\ast}_{1}(t_{1}-t_{2},j_1)\ldots\nonumber\\
     &&\prod\limits_{i_{n}=1}^{n}\mathcal{G}^{\ast}_{1}(t_{n}-t_{n},i_{n})
        \sum\limits_{k_{n}=1}^{n}\mathcal{N}^{\ast}_{\mathrm{int}}(k_{n},n+1)
        \prod\limits_{j_n=1}^{n+1}\mathcal{G}^{\ast}_{1}(t_{n},j_n)
        \otimes_{i=1}^{n+1}f_1^0(i)\otimes g_{n+1}^0(1,\ldots,n+1),\nonumber
\end{eqnarray}
and the limiting reduced density operators $f_n(t,1,\ldots,n),\,n\geq2,$ are determined by
the following expressions:
\begin{eqnarray}\label{dchaos}
     &&\hskip-7mm f_n(t,1,\ldots,n)=\prod_{i_1=1}^{n}\mathcal{G}^{\ast}_{1}(t,i_1)
       g_{n}^0(1,\ldots,n)\otimes \prod_{i_2=1}^{n}(\mathcal{G}_{1}^{\ast})^{-1}(t,i_2)
       \otimes_{j=1}^{n}f_{1}(t,j),\quad n\geq2,
\end{eqnarray}
where the inverse operator group to group \eqref{dgroup} is denoted by $(\mathcal{G}^{\ast}_{1})^{-1}(t)$.

Indeed, for the mean-value functionals of the limiting $k$-ary reduced observables \eqref{kIterd}
and the initial state specified by the one-particle density operator and the correlation operators
\eqref{isc}, the following representation holds on a finite time interval:
\begin{eqnarray*}
    &&\hskip-7mm \big(b^{(k)}(t),f^{(cc)}\big)=
       \frac{1}{k!}\mathrm{Tr}_{1,\ldots,k}\,b_{k}^{0}(1,\ldots,k)f_{k}(t,1,\ldots,k),\quad k\geq 2.
\end{eqnarray*}

For initial states specified by the sequence of limiting correlation operators \eqref{isc}, the process
of initial correlation propagation is described by the following sequence of correlation operators:
\begin{eqnarray*}
   &&\hskip-7mm g_n(t,1,\ldots,n)=\prod_{i_1=1}^{n}\mathcal{G}_{1}^{\ast}(t,i_1)
      \sum\limits_{\mbox{\scriptsize$\begin{array}{c}\mathrm{P}:(1,\ldots,n)=\bigcup_{i}X_{i}\end{array}$}}
      (-1)^{|\mathrm{P}|-1}(|\mathrm{P}|-1)!\,\otimes_{X_i\subset\mathrm{P}}g_{|X_i|}^0(X_{i})\otimes\\
   &&\prod_{i_2=1}^{n}(\mathcal{G}_{1}^{\ast})^{-1}(t,i_2)
      \otimes_{j=1}^{n}f_{1}(t,j), \quad n\geq2.
\end{eqnarray*}

The limiting one-particle density operator \eqref{vitercc} is governed by the quantum Vlasov kinetic
equation with initial correlations:
\begin{eqnarray}\label{Vls}
  &&\hskip-7mm \frac{\partial}{\partial t}f_{1}(t,1)=\mathcal{N}_1^{\ast}(1)f_{1}(t,1)+\\
  && \mathrm{Tr}_{2}\,\mathcal{N}^{\ast}_{\mathrm{int}}(1,2)
     \prod_{i_1=1}^{2}\mathcal{G}^{\ast}_{1}(t,i_1)g_{2}^0(1,2)\otimes
     \prod_{i_2=1}^{2}(\mathcal{G}^{\ast}_{1})^{-1}(t,i_2)f_{1}(t,1)\otimes f_{1}(t,2),\nonumber\\
\label{Vlasovic}
  &&\hskip-7mm f_{1}(t,1)|_{t=0}=f_{1}^0(1),
\end{eqnarray}
where the notation adopted in equation \eqref{Vlasov1} is used.

It should be noted that the resulting kinetic equation \eqref{Vls} is non-Markovian. For pure states,
kinetic equation \eqref{Vls} reduces to the Hartree equation with initial correlations. In the case
of a separable initial state \eqref{h}, kinetic equation \eqref{Vls} coincides with the quantum Vlasov
kinetic equation \eqref{Vlasov1}, and the reduced density operators \eqref{dchaos} describe the propagation
of initial chaos \eqref{ihaos}.

We also note that, for arbitrary initial states in the mean-field limit, the sequence of limiting reduced 
density operators, as well as the sequence \eqref{vitercc},\eqref{dchaos}, is a solution to the Cauchy 
problem for the Vlasov hierarchy:
\begin{eqnarray*}
     &&\hskip-7mm \frac{\partial}{\partial t}f_{s}(t,1,\ldots,s)=
        \sum\limits_{i=1}^{s}\mathcal{N}^{\ast}_1(i)f_{s}(t,1,\ldots,s)+\\
     &&\sum\limits_{i=1}^{s}\mathrm{Tr}_{s+1}\mathcal{N}^{\ast}_{\mathrm{int}}(i,s+1)f_{s+1}(t,1,\ldots,s+1),\\
     &&\hskip-7mm f_{s}(t,1,\ldots,s)|_{t=0}=f_{s}^{0}(1,\ldots,s),\quad s\geq1,
\end{eqnarray*}
where the operators $\mathcal{N}^{\ast}_{1}$ and $\mathcal{N}^{\ast}_{\mathrm{int}}$ are defined according
 to formulas \eqref{vNeg}.
This hierarchy of evolution equations is dual to the hierarchy \eqref{vdh} in the sense of the mean-value
functional for the limiting reduced observables \eqref{Iterd}.

Thus, for initial states specified by a one-particle density operator and correlation operators \eqref{isc},
the dual Vlasov hierarchy \eqref{vdh} for the limiting reduced additive-type observables describes the 
evolution of quantum many-particle systems, just as the non-Markovian Vlasov quantum kinetic equation with 
the initial correlations \eqref{Vls} does. In the case of limiting reduced nonadditive-type observables, the 
dual Vlasov hierarchy \eqref{vdh} describes equivalently, in the sense of equality \eqref{dchaos}, the 
propagation of initial correlations in terms of reduced density operators \eqref{dchaos}. In other words, 
the alternative method of describing the evolution of the state of quantum systems in the mean-field 
approximation is based on the non-Markovian Vlasov kinetic equation with initial correlations \eqref{Vls}.
The results can be extended to many-body bosonic or fermionic systems, as in \cite{GP11}, to study the 
dynamics of correlations.

\textcolor{blue!50!black}{\section{Conclusion and outlook}}

This article examines the evolution equations of quantum many-particle systems and, in essence, classifies
them. Hierarchies of such evolution equations describe the evolution of observables and the state of systems
with both finite and infinite numbers of particles, which underlie the description of kinetic evolution.

The article introduces the concepts of cluster \eqref{cexp},\eqref{dcexp} and cumulant \eqref{rocum},\eqref{drcum}
expansions of operator groups, which underlie the description of the evolution of observables \eqref{sroh} and the
state \eqref{sdh}, respectively, in quantum many-particle systems. The generating operators for the non-perturbative
representation of solutions to the hierarchies of evolution equations for the state \eqref{drcum} and for the
hierarchy of evolution equations for observables \eqref{sroh} are determined, respectively, by the cumulants
(semi-invariants) of dual operator groups \eqref{dgroup} and the cumulants of operator groups \eqref{group}
for quantum systems with a finite number of particles.

Non-perturbative solutions to the Cauchy problem for hierarchies of evolution equations for the state have
clarified their relationship \eqref{Fg} and \eqref{Gg} to the dynamics of correlations \eqref{vNhs} in quantum
particle systems. In passing, laws \eqref{fch} and \eqref{cch}  governing processes of correlation creation from
chaos were established.

The notion of cumulants of operator groups also underlies the kinetic description of the collective behavior
of quantum systems with an infinite number of particles. The methods for deriving kinetic equations using
non-perturbative solutions of hierarchies of evolution equations of quantum many-particle systems are based
on the notion of kinetic cluster expansions \eqref{kce} of cumulants of operator groups.

For separable initial states \eqref{si}, we established the equivalence of the Cauchy problems for the
hierarchies of evolution equations for reduced density operators and reduced observables to the Cauchy
problem for the generalized quantum kinetic equation \eqref{gke} and a sequence of correlation functionals
\eqref{rf} describing the propagation of correlations in quantum many-particle systems. Proving this result
involves applying kinetic cluster expansions \eqref{kce} to the generating operators of a non-perturbative
representation of solutions of these hierarchies of evolution equations.

This article also addresses the challenge of describing kinetic evolution in terms of observables \eqref{vdh},
in contrast to the conventional approach based on describing the effective evolution of the state of many quantum
particles. One of the advantages of the developed approach is the opportunity to construct kinetic equations
\eqref{Vls} that account for particle correlations in the initial state, as well as to describe the propagation
of initial correlations \eqref{dchaos} within scaling approximations.

An alternative perspective on the nature of kinetic equations based on describing the evolution of correlations
in systems of many quantum particles is also presented. The approach describes the evolution of a state through
the von Neumann hierarchy of evolution equations \eqref{vNh} for a sequence of cumulants of the density
operators \eqref{gD}.

\bigskip

\textbf{Acknowledgments.} \small{The key findings of this article were discussed during the Year of Quantum Science.
The author expresses heartfelt gratitude to the organizers of StatPhys29 for the inspiring conference atmosphere
and financial support during my visit to Florence. The author also thanks the Isaac Newton Institute (INI) and
the London Mathematical Society (LMS)  for their financial support, which made essential workshops and meetings
possible, as well as the School of Mathematics and Physics at the University of Surrey for their warm hospitality.}

\bigskip

\small{

}
\end{document}